\documentclass[11pt, oneside, letter]{scrartcl}
\usepackage[english]{babel}
\usepackage[utf8]{inputenc}
\usepackage{color}
\usepackage{geometry}
\usepackage{amsmath}
\usepackage{amssymb}
\usepackage{amsthm}
\usepackage{enumitem}
\usepackage{mathtools}
\usepackage{abraces}
\usepackage{dsfont}
\usepackage{colonequals}
\usepackage{mathrsfs}
\usepackage{nicefrac}
\usepackage{siunitx}
\usepackage{hyperref}

	\RequirePackage[]{geometry}
	\usepackage[
		headtopline, 
		headsepline, 
		footsepline, 
		footbotline  
		]{scrlayer-scrpage}
		\clearpairofpagestyles
		
		\automark[section]{section} 
		\ihead{\headmark}
		\ohead{\shortname}
	\newcommand{\shortname}{Decay of Entropy and Information}

	\theoremstyle{definition} 
	\newtheorem{defi}{Definition}[section]
	\newtheorem*{defi*}{Definition}
	
	\newtheorem*{nota*}{Notation}
	
	\newtheorem*{bsp*}{Beispiel}
	
	\newtheorem*{bem*}{Bemerkung}
	\newtheorem*{anm*}{Anmerkung}
	
	\theoremstyle{theorem} 
	\newtheorem{theo}[defi]{Theorem}
	\newtheorem*{theo*}{Theorem}
	
	\newtheorem*{satz*}{Satz}
	\newtheorem{lem}[defi]{Lemma}
	\newtheorem*{lem*}{Lemma}
	
	\newtheorem*{kor*}{Korollar}
	\newtheorem*{abs*}{Abstract}

	\newtheorem*{prop*}{Proposition}
	
	\newtheorem{cla*}{Claim}
	\newtheorem{cor}[defi]{Corollary}
	\newtheorem{cor*}{Corollary}
	
	\theoremstyle{definition}	
	\newtheorem{rem}[defi]{Remark}
	\newtheorem*{rem*}{Remark}
	
	\newtheorem*{exa*}{Example}
	\newtheorem*{rec*}{Recall}

		\newcommand{\calI}{\mathcal{I}}

		\newcommand{\calL}{\mathcal{L}}
		\newcommand{\calM}{\mathcal{M}}
	\newcommand{\bbN}{\mathbb{N}}	
		
		\newcommand{\calP}{\mathcal{P}}
		
	\newcommand{\bbR}{\mathbb{R}}	
	\newcommand{\bbS}{\mathbb{S}}	
		
		\newcommand{\calU}{\mathcal{U}}

	\newcommand{\bbone}{\mathds{1}}

	\DeclareMathOperator{\lscal}{\langle}		
	\DeclareMathOperator{\rscal}{\rangle}		

	\definecolor{DefColor}{rgb}{0.13, 0.60, 0.13}

	\newcommand\quotient[2]{
		\mathchoice	{\textup{\raise1ex\hbox{$#1$}\Big/\lower1ex\hbox{$#2$}}}{#1\,/\,#2}{#1\,/\,#2}{#1\,/\,#2}}

\newcommand{\boldu}{\boldsymbol{u}}
\newcommand{\boldv}{\boldsymbol{v}}
\newcommand{\boldw}{\boldsymbol{w}}
\newcommand{\boldx}{\boldsymbol{x}}
\newcommand{\boldy}{\boldsymbol{y}}
\newcommand{\boldp}{\boldsymbol{p}}
\newcommand{\boldq}{\boldsymbol{q}}
\newcommand{\boldA}{\boldsymbol{A}}
\newcommand{\boldB}{\boldsymbol{B}}
\newcommand{\boldC}{\boldsymbol{C}}
\newcommand{\boldD}{\boldsymbol{D}}
\newcommand{\boldK}{\boldsymbol{K}}
\newcommand{\boldL}{\boldsymbol{L}}
\newcommand{\boldM}{\boldsymbol{M}}
\newcommand{\boldN}{\boldsymbol{N}}
\newcommand{\boldP}{\boldsymbol{P}}
\newcommand{\balpha}{\boldsymbol{\alpha}}
\newcommand{\bsigma}{\boldsymbol{\sigma}}

\DeclareMathOperator{\slim}{s-lim}		

\newcommand{\di}{d}					
\newcommand{\numkac}{N} 			
\newcommand{\numhr}{M}				
\newcommand{\Ekin}{K} 				
\newcommand{\Ekinkac}{K_S}			
\newcommand{\Ekinhr}{K_R}			
\newcommand{\Mom}{P} 				
\newcommand{\Proj}{\boldP}			

\newcommand{\SG}{g} 			
\newcommand{\TS}{h}				
\newcommand{\TSone}{\TS^{(1)}}
\newcommand{\TStwo}{\TS^{(2)}}
\newcommand{\OUone}{P_s^{(1)}}
\newcommand{\OUtwo}{P_s^{(2)}}

\newcommand{\ColMat}{M_\sigma}
\newcommand{\ColMatij}{\boldsymbol{M}_\sigma^{(i,j)}}
\newcommand{\NCol}{N_\sigma}
\newcommand{\NColi}{\boldN_\sigma^{(i)}}
\newcommand{\NColj}{\boldN_\sigma^{(j)}}

\newcommand{\wColj}{\boldw_\sigma^{(j)}}
\newcommand{\pmea}{\rho}

\newcommand{\MargG}{\calM}		
\newcommand{\MwS}{g}			
\newcommand{\MwN}{g}			
\newcommand{\testH}{h}			
\newcommand{\intbbS}{\int_{\bbS^{\di-1}}}
\newcommand{\intkbbS}{\int_{\bbS^{k(\di-1)}}}
\newcommand{\intrt}{\int_{\bbR^\di}}
\newcommand{\intrtN}{\int_{\bbR^{\di\numkac}}}
\newcommand{\intrtM}{\int_{\bbR^{\di\numhr}}}
\newcommand{\intrtNM}{\int_{\bbR^{\di(\numkac+\numhr)}}}
\newcommand{\LOneKac}{L^1\left(\bbR^{\di\numkac},\textup{d}\boldv\right)}
\newcommand{\HOneKac}{H^1\left(\bbR^{\di\numkac},\textup{d}\boldv\right)}
\newcommand{\LOneGauss}{L^1\left(\bbR^{\di\numkac}, \MwS(\boldv)\textup{d}\boldv\right)}
\newcommand{\HOneGauss}{H^1\left(\bbR^{\di\numkac}, \MwS(\boldv)\textup{d}\boldv\right)}
\newcommand{\LOneKNM}{L^1\left(\bbR^{\di(\numkac+\numhr)},\textup{d}\boldu\right)}

\newcommand{\calLTS}{\calL'}
\newcommand{\Ent}{S}
\newcommand{\OUalpha}{\alpha}
\newcommand{\OUbeta}{\delta}
\newcommand{\invtemp}{\beta}
\newcommand{\invhr}{\beta}

\DeclareMathOperator{\InfK}{\calI} 
\DeclareMathOperator{\InfTS}{\tilde{\calI}} 
\DeclareMathOperator{\diag}{diag} 

	\def\XXint#1#2#3{{\setbox0=\hbox{$#1{#2#3}{\int}$ }
			\vcenter{\hbox{$#2#3$ }}\kern-.6\wd0}}

	\newcommand\numberthis{\addtocounter{equation}{1}\tag{\theequation}}
	
	\setkomafont{subparagraph}{\normalfont\itshape} 
	\RedeclareSectionCommand[beforeskip=-0.5\baselineskip]{subparagraph}

\begin{document}
		
	\thispagestyle{empty}
	
	\begin{center}
		\Large
		\textbf{
		Decay of Entropy and Information\\for multidimensional Kac models}
		
		\large
		
		Lukas Hauger\footnote{
		Max-Planck-Institut for Mathematics in the Sciences, Leipzig, Germany (lukas.hauger@mis.mpg.de). The paper was created at the School of Mathematics of the Georgia Institute of Technology, Atlanta, United States, as part of a master's thesis under the supervision of Michael Loss .
	}
	\vspace{-9pt}
	\end{center}
	\normalsize
		
	\paragraph{Abstract}
	
	\textit{We study the approach to equilibrium of systems of gas particles in terms of relative entropy. The systems are modeled by the Kac master equation in arbitrary dimensions. First, we study the Kac system coupled to a thermostat, and secondly connected to a heat reservoir. The use of the Fisher-information allows elementary proofs with weak regularity assumptions. As a result, we obtain for both systems exponential decay rates for the entropy and information that are essentially independent of the size of the systems.}
	
	\vspace{-12pt}
	
	\tableofcontents

\section{Introduction}

	In the seminal paper \cite{KacFoundations} from 1956, Mark Kac suggested a probabilistic model to describe a spatially homogeneous system of gas particles, now called the \textit{Kac Model}. The velocities $v_1,\dots,v_\numkac\in\bbR^\di$ of the $\numkac\in\bbN$ particles are combined into a "master vector" $\boldv=(v_1,\dots,v_\numkac)$ that describes the state of the system. Energy and momentum preserving pair collisions are introduced by the transformation of the master vector. The occurrence of the collisions in time is modeled by a Poisson-like process. This leads to a linear evolution equation for the probability density function describing the system, called the \textit{Kac master equation}. For details about the assumptions of the Kac model, and the derivation of the Kac master equation from them, see \cite{KacFoundations} or \cite{ManyBodyKac}[Chapter 2, p. XI-5].
	
	The importance of the Kac model for us is due to the following two reasons: First, the non-linear spatially homogeneous Boltzmann equation can be derived from the Kac model via the notion of propagation of chaos, see \cite{KacFoundations}. Second, the Kac model enables the investigation of the approach to equilibrium in relative entropy.
	
	A natural measure for the approach to equilibrium is the relative entropy of the system. The relative entropy is an extensive quantity, i.e.\ it is proportional to the size of the system. We would like to obtain an exponential decay rate that is essentially independent of the size of the system. However, the results of Cédric Villani and Amit Einav (see \cite{VillaniEntDecArbIC} and \cite{EinavExEntDecArbIC}) suggest that this is not possible for arbitrary initial conditions. Therefore, it is necessary to consider specific classes of initial conditions.
	
	In this paper, we consider a multidimensional Kac system first coupled to a thermostat, and secondly coupled to a heat reservoir. We obtain exponential decay rates for the entropy and information of the Kac system in both models that are essentially independent of the size of the systems. The proofs for the decay of information are elementary and require only weak assumptions. The obtained decay rates for the information are then transferred to the entropy. This information approach was first used in \cite{DecOfInfKac} for a one-dimensional Kac model coupled to a heat reservoir.
	
	The thermostat is treated as an infinite gas at thermal equilibrium. Thus, it always remains in its equilibrium state. In \cite{KacThermostat}, the entropy decay of a one-dimensional Kac model coupled to a thermostat was already derived. However, the proof does not carry over to the multidimensional case due to technical difficulties. The approach via information used in this paper solves that problem.
	
	The heat reservoir is assumed to be much larger than the Kac system and is only initially in a thermal equilibrium. It is therefore a more realistic version of the thermostat. The fact, that one does not know how the heat reservoir evolves in time makes this model more difficult.
	
	\paragraph{Acknowledgments}
	
	I sincerely thank my supervisor, Michael Loss, whose support and guidance made this work possible. Thank you for introducing me to the fascinating topic of kinetic gas theory. I am also thankful to the School of Mathematics at the Georgia Institute of Technology for providing excellent research conditions.

\section{The Kac system coupled to a thermostat}

	In this section, we will first lay out the model, apply a ground state transformation, and then state the main results.

	\subsection{Model}

	We combine the $\di$-dimensional velocities of the $\numkac$ particles in the Kac system to a master vector $\boldv = (v_1,\dots,v_\numkac)\in\bbR^{\di\numkac}$.	The state of the Kac system at time $t\geq0$ is described by a probability density function $f_t\in\LOneKac$. The time evolution is governed by the linear Kac master equation
	\begin{equation}
		\label{EqMasterEquation}
		\frac{\textup{d}}{\textup{d} t} f_t
		= \calL f_t.
	\end{equation}
	The time evolution operator $\calL$ is defined by
	\begin{equation}
		\calL
		\coloneqq \lambda \numkac (Q-\bbone) + \mu \numkac (R-\bbone).
	\end{equation}
	Hereby, the collision operator $Q$ representing the collisions between particles within the Kac system is given by\footnote{
		For $\numkac=1$ we set $Q=\bbone$ so that $\bbone-Q = 0$, and the first part in the master equation vanishes. Note that this follows naturally, if we define the first part of the master equation via $(\bbone-Q) \coloneqq \binom{\numkac}{2}^{-1} \sum_{i<j} \big(\bbone-\tilde{Q}_{i,j}\big)[f]$, where $\tilde{Q}_{i,j}[f](\boldv) \coloneqq \intbbS f\big(\ColMatij \boldv,t\big) \textup{d}\pmea(\sigma)$.}
	\begin{equation}
		Q[f](\boldv)
		\coloneqq \binom{\numkac}{2}^{-1} \sum_{1\leq i < j \leq \numkac} \intbbS f\left(\ColMatij \boldv \right) \,\textup{d}\rho(\sigma).
	\end{equation}
	We use the reflection map to model the collisions between two particles $v,w\in\bbR^\di$ with scattering angle $\sigma \in \bbS^{\di-1}$
	\begin{equation}
		\label{EqReflectionMap}
		\begin{pmatrix} v \\ w \end{pmatrix}
		\mapsto \ColMat \begin{pmatrix} v \\ w \end{pmatrix}
		= \begin{pmatrix}
		v - (\sigma[v-w])\sigma \\
		w + (\sigma[v-w])\sigma
		\end{pmatrix},
		\quad
		\ColMat
		\coloneqq \begin{pmatrix} \bbone-\sigma\otimes\sigma & \sigma\otimes\sigma \\ \sigma\otimes\sigma & \bbone-\sigma\otimes\sigma
		\end{pmatrix}.
	\end{equation}
	Note that the reflection map only depends on the tensor product $\sigma\otimes\sigma$ of the scattering angle.
	
	The kinematic of the collisions is explained in great detail in \cite[Appendix A]{StrConMaxMod}. The collision mapping is clearly kinetic energy and momentum conserving. The collision between the $i$-th and $j$-th particle is represented by the collision matrix $\ColMatij \in \left(\bbR^{\di\times\di}\right)^{\numkac\times\numkac}$ defined by
	\begin{equation}
		\label{EqDefColMatij}
		\ColMatij
		\coloneqq \begin{pmatrix}
		\bbone & 0 & \cdots & & & \cdots & 0 \\
		0 & \ddots & & & & & \vdots \\
		\vdots & & \bbone- \sigma\otimes\sigma & \cdots & \sigma\otimes\sigma & & \\
		& & \vdots & & \vdots & & \\
		& & \sigma\otimes\sigma & \cdots & \bbone- \sigma\otimes\sigma & & \vdots \\
		\vdots & & & & & \ddots & 0 \\
		0 & \cdots & & & \cdots & 0 & \bbone
		\end{pmatrix}.
	\end{equation}
	Note that $\ColMatij$ is the identity matrix except for the four entries $(i,i)$, $(i,j)$, $(j,i)$, and $(j,j)$ that are substituted by the entries of $\ColMat$. In other words, $\ColMatij \boldv	= (v_1,...,v_i^*,...,v_j^*,...,v_\numkac)$ with $(v_i^*,v_j^*)^T = \ColMat (v_i,v_j)^T$.
	
	For each collision, a scattering angle is selected by a probability measure $\rho$ on the sphere $\bbS^{\di-1}$, for which we assume the symmetry condition\footnote{In fact, the results in this paper depend only on the symmetry condition and not on the probability measure itself. We thank Felix Otto for this useful remark.}
	\begin{equation}
		\label{EqPmeaSym}
		\intbbS \sigma\otimes\sigma \,\textup{d}\pmea(\sigma)
		= \frac{1}{\di}\bbone_\di.
	\end{equation}
	The thermostat is assumed to contain infinitly many particles. Therefore, the velocity distribution of the particles in the thermostat is not influenced by the collision process, i.e.\ the distribution of the particles in the thermostat remains constant over time. This leads to the following operator $R:\LOneKac\to\LOneKac$ representing the interaction with the thermostat
	\begin{equation}
		R
		\coloneqq \frac{1}{\numkac} \sum_{j=1}^\numkac R_j,
	\end{equation}
	with $R_j:\LOneKac\to\LOneKac$ defined by
	\begin{equation}
		R_j[f]
		\coloneqq \intrt \textup{d}w\, \intbbS \textup{d}\pmea(\sigma)\,
		\MwS\left( \left[ \boldM_\sigma^{(j,\numkac+1)} \begin{pmatrix} \boldv \\ w \end{pmatrix} \right]_{\numkac+1} \right)
		f\left( \Proj \boldM_\sigma^{(j,\numkac+1)} \begin{pmatrix} \boldv \\ w \end{pmatrix} \right).
	\end{equation}
	Hereby, $\Proj: \bbR^{\numkac+1} \to \bbR^\numkac, (v_1,\dots,v_\numkac,w) \mapsto (v_1,\dots,v_\numkac)$ denotes a projection onto the first $\numkac$ variables.
	
	The parameter $\frac{1}{\lambda}$ describes the average waiting time for a particle in the Kac system to collide with any particle in the Kac system. Similarly, the parameter $\frac{1}{\mu}$ gives the average waiting time for a particle in the Kac system to collide with any particle in the thermostat.
	
	Note that the collision operators $Q$ and $R_j$ only depend on the scattering angle $\sigma$ by the integral over a function of the tensor product $\sigma\otimes\sigma$. Therefore, the process describing our model is entirely defined by the symmetry condition \eqref{EqPmeaSym}, i.e.\ all probability measures that fulfill the symmetry condition generate the same process.
	
	Finally, we assume that the velocity vectors of the particles in the thermostat are distributed according to the Maxwell distribution
	\begin{equation}
		\label{EqDefMwN}
		\MwN(\boldv)
		\coloneqq \left( \frac{\invtemp}{2\pi} \right)^\frac{\di\numkac}{2} e^{-\frac{\invtemp}{2}\boldv^2},
		\quad \boldv \in \bbR^{\di\numkac}.
	\end{equation}
	Hereby, the mass is assumed to be $m=1$ and the inverse temperature is $\invtemp \coloneqq \frac{1}{kT}$.

	\subsection{Kinetic energy, entropy and Fisher-information}

	A first observation is that the time evolution of the total kinetic energy and the total momentum can be computed explicitly, as done in \cite[p. 6]{KacThermostat} for a one-dimenional model.
	
	\begin{defi*}
		For any solution $(f_t)_{t\geq0}$ of the Kac master equation \eqref{EqMasterEquation} with $f_0\in \LOneKac$, the \textit{total kinetic energy} $\Ekin$ \textit{of the Kac system at time} $t\in[0,\infty)$ is defined by
		\begin{equation}
			\label{EqDefEkinThermo}
			\Ekin(t)
			\coloneqq \frac{1}{2} \intrtN \boldv^2 f_t(\boldv)  \,\textup{d}\boldv.
		\end{equation}
		The \textit{total momentum} $\Mom(t)$ \textit{of the Kac system} is defined by
		\begin{equation}
			\label{EqDefMomThermo}
			\Mom(t)
			\coloneqq \sum_{i=1}^\numkac \intrtN v_i f_t(\boldv) \,\textup{d}\boldv.
		\end{equation}	
	\end{defi*}
	Differentiating under the integral sign, and inserting the master equation \eqref{EqMasterEquation} leads to a first-order linear inhomogeneous ODE, from which we obtain the following lemma.
	
	\begin{lem}
		\label{LemEkinMom}
		The total kinetic energy $\Ekin(t)$ of the Kac system for any solution $(f_t)_{t\geq0}$ of the Kac master equation \eqref{EqMasterEquation} with $f_0 \in \LOneKac$ and finite initial kinetic energy, i.e.\ $\Ekin(0)<\infty$, is given by
		\begin{equation}
		\label{EqEkinThermostat}
		\Ekin(t)
		= \left( \Ekin(0) - \frac{\di\numkac}{\invtemp} \right) e^{-\frac{\mu}{2\di} t} + \frac{\di\numkac}{\invtemp}.
		\end{equation}
		The total momentum $\Mom(t)$ of the Kac system for any solution $(f_t)_{t\geq0}$ of the Kac master equation \eqref{EqMasterEquation} with $f_0 \in \LOneKac$ and finite initial momentum, i.e.\ $\Mom(0)<\infty$, is given by
		\begin{equation}
		\Mom(t)
		= e^{-\frac{\mu}{\di}t} \Mom(0).
		\end{equation}
	\end{lem}
	Lemma \ref{LemEkinMom} shows that the average kinetic energy of a particle in the Kac system tends exponentially towards the average kinetic energy of a particle in the thermostat. It is interesting to note that \eqref{EqEkinThermostat} is independent of the rate parameter $\lambda$, i.e.\ the collisions between particles in the Kac system do not influence the time evolution of the kinetic energy of the Kac system. This explains that the rate of convergence is independent of the number of particles in the Kac system. Similarly, the total momentum tends exponentially to zero, and is not influenced by the collisions between particles in the Kac system.
	
	\begin{defi*}
		We define the \textit{relative entropy} of a state $f_t \in \LOneKac$ at time $t \in [0,\infty)$ \textit{with respect to the equilibrium state} $\gamma$ as (see \cite[p. 4]{KacThermostat})
		\begin{equation}
			\label{EqDefEntropy}
			\Ent(f_t|\gamma)
			\coloneqq \intrtN f_t(\boldv) \ln\left( \frac{f_t(\boldv)}{\gamma(\boldv)} \right) \,\textup{d}\boldv.
		\end{equation}
		Note that the entropy is defined with the inverted sign. 
	\end{defi*}
	
	Before turning to the information, it is convenient to write our model with the constant function $H\equiv1$ as ground state instead of the Gaussian $\MwN$, see \cite[p. 14]{KacThermostat}. For a solution $(f_t)_{t\geq0}$ of the master equation \eqref{EqMasterEquation} we define $(\TS_t)_{t\geq 0}\in\LOneGauss$ by
	\begin{equation}
		\label{EqGSTransformation}
		\TS_t(\boldv)
		\coloneqq \frac{f_t(\boldv)}{\SG(\boldv)}.
	\end{equation}
	Inserting the transformed PDF into the Kac master equation \eqref{EqMasterEquation} yields the following transformed master equation.
	\begin{lem}
		\label{LemDerivativeTS}
		The functions $(f_t)_{t\geq0}\in\LOneKac$ form a solution of the Kac master equation \eqref{EqMasterEquation} if and only if $(\TS_t)_{t\geq0} \in \LOneGauss$ defined by \eqref{EqGSTransformation} solves the transformed master equation
		\begin{equation}
			\label{EqMasterEquationTS}
			\frac{\partial \TS_t}{\partial t}
			= \calLTS[\TS_t]
			\coloneqq \lambda \numkac (Q-\bbone)[\TS_t] + \mu \numkac (T-\bbone)[\TS_t],
		\end{equation}
		where $T \coloneqq \frac{1}{\numkac} \sum_{j=1}^\numkac T_j$ with
		\begin{equation}
			T_j[\TS_t](\boldv)
			\coloneqq \intrt \MwS(w)\textup{d}w \intbbS \textup{d}\pmea(\sigma)\,
			\TS_t\left( \Proj \boldM_\sigma^{(j,\numkac+1)} (\boldv, w)^T \right).
		\end{equation}
	\end{lem}
	
	\begin{defi*}
		We define the Fisher-information with respect to the Gaussian measure $\MwN(\boldv)\textup{d}\boldv$ of a probability density function $\TS:\bbR^{\di\numkac}\to\bbR$ with $\sqrt{\TS}\in\HOneGauss$ by
		\begin{equation}
			\label{EqDefInfTS}
			\InfTS(\TS)
			\coloneqq \intrtN \frac{|\nabla \TS(\boldv)|^2}{\TS(\boldv)} \,\MwN(\boldv)\textup{d}\boldv.
		\end{equation}
	\end{defi*}

	\subsection{Main results}

	Our first main result is the decay of the information of the Kac system.
	\begin{theo}
		\label{TheoDecayInf}
		Let $(\TS_t)_{t\geq0}$ be a solution of the transformed master equation \eqref{EqMasterEquationTS} with the symmetry condition \eqref{EqPmeaSym} such that $\sqrt{\TS_0}\in \HOneGauss$. Then, the information decays exponentially in time with
		\begin{equation}
			\InfTS(\TS_t)
			\leq e^{-\frac{1}{\di} \mu t} \InfTS(\TS_0).
		\end{equation}
		The proof is given in section \ref{ProofTheoDecayInf}.
	\end{theo}

	The connection between information and entropy is established via the Ornstein-Uhlenbeck semigroup $(P_s)_{s\geq0}$ (see \cite[p.4, eq. (12)]{DecOfInfKac})
	\begin{equation}
		\Ent(f|\MwN)
		= \frac{1}{\invtemp} \int_0^\infty \InfTS\left(P_s e^{\calL' t} \TS_0 \right) \,\textup{d}s.
	\end{equation}
	Recall the definition of the Ornstein-Uhlenbeck semigroup, see appendix \ref{DefiOUsemigroup}.
	
	This raises the question, if the Ornstein-Uhlenbeck semigroup commutes with the time evolution operator $\calLTS$, since a decay rate for the information
	\begin{equation}
		\InfTS(\TS_t)
		\leq c(t) \InfTS(\TS_0),
	\end{equation}
	would then imply the same decay rate for the entropy:
	\begin{align*}
		\Ent(f_t|\MwS)
		&= \frac{1}{\invtemp} \int_0^\infty \textup{d}s \InfTS\left(P_se^{t\calL'} \TS_0\right)
		= \frac{1}{\invtemp} \int_0^\infty \textup{d}s \InfTS\left(e^{t\calL'} P_s \TS_0\right) \\
		&\leq \frac{1}{\invtemp} c(t) \int_0^\infty \textup{d}s \InfTS(P_s \TS_0)
		= c(t) \Ent(f_0|\MwS).
		\label{EnEntEstByInf} \numberthis
	\end{align*}
	
	For the reflection map, that we use as collision mechanism, the Ornstein-Uhlenbeck semigroup indeed commutes with the time evolution operator $\calLTS$.
	
	\begin{theo}
		\label{TheoCommOUTimeEvolution}
		The Ornstein-Uhlenbeck semigroup commutes with the time evolution operator $\calL'$, i.e.\ for all $s\geq0$ we have
		\begin{equation}
			P_s\calL' = \calL' P_s.
		\end{equation}
		Hereby, the operators are assumed to act on $L^1\left(\bbR^{\di\numkac},\MwN(\boldv)\textup{d}\boldv\right)$. The proof is given in section \ref{ProofTheoCommOUTimeEvolution}.
	\end{theo}
	
	Therefore, theorem \ref{TheoCommOUTimeEvolution} implies by inequality \eqref{EnEntEstByInf} the following theorem about the transfer of the decay rate from the information to the entropy.
	
	\begin{theo}
		\label{TheoEntEstByInfEst}
		Let $(f_t)_{t\geq0}$ be a solution of the Kac master equation \eqref{EqMasterEquation} with the symmetry condition \eqref{EqPmeaSym} such that $\sqrt{f_0} \in \HOneGauss$ and $\Ekin(0)<\infty$. Assume we have for $(\TS_t)_{t\geq0}$ defined by \eqref{EqGSTransformation} an inequality for the information
		\begin{equation}
			\label{EqTheoEntEstByInfEstInfEst}
			\InfTS(\TS_t)
			\leq c(t) \InfTS(\TS_0).
		\end{equation}
		Then, we get an inequality for the relative entropy by
		\begin{equation}
			\Ent(f_t|\MwS)
			\leq c(t) \Ent(f_0|\MwS).
		\end{equation}
	\end{theo}	
	
	Theorem \ref{TheoDecayInf} together with theorem \ref{TheoEntEstByInfEst} prove the entropy decay as stated in the following theorem.
	
	\begin{theo}
		\label{TheoDecayEntropy}
		Let $(f_t)_{t\geq0}$ be a solution of the master equation \eqref{EqMasterEquation} with the symmetry condition \eqref{EqPmeaSym} such that $\sqrt{f_0} \in \HOneKac$ and $\Ekin(0)<\infty$. Then, the relative entropy decays exponentially with
		\begin{equation}	
			\Ent(f_t|\MwS)
			\leq e^{-\frac{1}{\di} \mu t}\Ent(f_0|\MwS).
		\end{equation}
	\end{theo}


\section{The Kac system coupled to a heat reservoir}

	In this section, we replace the thermostat by a finite heat reservoir. The main difference is that the number of particles in the heat reservoir $\numhr \in\bbN$ is assumed to be finite. Thus, the state of the heat reservoir changes over time.
	
	\subsection{Model}
	
	The system is described by a probability distribution $F:\bbR^{\di\numkac}\times\bbR^{\di\numhr}\to\bbR$ of the velocity vectors of all particles, normalized with respect to the Lebesgue measure.	We use the notation $(\boldv,\boldw) = (v_1,\dots,v_\numkac,w_{\numkac+1},\dots,w_{\numkac+\numhr})$ to number the particles from $1$ to $\numkac+\numhr$. The particles $1,\dots,\numkac$ belong to the Kac system, and the particles $\numkac+1,\dots,\numkac+\numhr$ to the heat reservoir. The pair collisions are modeled by the operator $R_{ij}: \LOneKNM\to\LOneKNM$ given by
	\begin{equation}
		R_{ij}[F](\boldu)
		\coloneqq \intbbS \textup{d}\rho(\sigma)\, F\left( \ColMatij \boldu \right),
		\quad \boldu \in \bbR^{\di(\numkac+\numhr)},
	\end{equation}
	where we model the collision between the $i$-th and $j$-th particle again by the reflection map, i.e.\ $\ColMatij \in \bbR^{(\numkac+\numhr)\times(\numkac+\numhr)}$ is defined as in \eqref{EqDefColMatij} and $\pmea$ is a probability measure fulfilling the symmetry condition \eqref{EqPmeaSym}. The time evolution operator $\calL$ is given by (compare to \cite[p. 2]{DecOfInfKac})
	\begin{equation}
	\label{EqDefCalLHR}
	\calL
	\coloneqq
	\frac{\lambda_S}{\numkac-1} \sum_{1\leq i < j \leq \numkac} (R_{ij}-\bbone)
	+ \frac{\lambda_R}{\numhr-1} \sum_{\numkac<i<j\leq \numkac+\numhr} (R_{ij}-\bbone)
	+ \frac{\mu}{\numhr} \sum_{i=1}^\numkac \sum_{j=\numkac+1}^{\numkac+\numhr} (R_{ij}-\bbone).
	\end{equation}
	
	The parameter $\frac{1}{\lambda_S}$ describes the average waiting time for some particle in the Kac system to collide with any other particle in the Kac system. Analogously, the parameter $\lambda_R$ gives the rate of collisions within the heat reservoir. Finally, the rate at which a particle of the Kac system collides with any particle of the heat reservoir is $\mu$.
	
	The time evolution operator $\calL$ generates the exponential semigroup $\left(e^{\calL t}\right)_{t\geq0}$. The time evolution of the system is given by
	\begin{equation}
		\label{EqMasterEqHeatResCauchy}
		F_t = e^{\calL t}F_0.
	\end{equation}
	This is equivalent to the following abstract Cauchy problem, called the \textit{Kac Master Equation}
	\begin{equation}
		\label{EqMasterEqHeatRes}
		\frac{\partial F_t}{\partial t}
		= \calL F_t.
	\end{equation}
	Our goal is to investigate the approach to equilibrium of the Kac system. We assume, that the heat reservoir is initially in an equilibrium state. As in \cite[p. 2]{DecOfInfKac}, the initial probability distribution of the heat reservoir is therefore assumed to be a Gaussian function. Hence, we can write
	\begin{equation}
		\label{EqInitalStateF_0}
		F_0(\boldv,\boldw)
		= f_0(\boldv) \underbrace{\left(\frac{\invhr}{2\pi}\right)^\frac{\di\numhr}{2} e^{-\frac{\invhr}{2} \boldw^2}}_{\eqqcolon \MwN(\boldw)}
		= f_0(\boldv) \MwN(\boldw),
	\end{equation}
	where $\invhr$ denotes the inverse temperature of the initial state of the heat reservoir.

	\subsection{Basic notions}

	We are interested in the time evolution of the probability distribution $f_t$ describing the Kac system. We define $f_t$ by the marginal of $F_t$ with respect to the Lebesgue measure over the particles in the heat reservoir (see \cite[eq. (6)]{DecOfInfKac} or \cite[eq. (8)]{EntDecKac})
	\begin{equation}
		\label{EqDeff_t}
		f_t(\boldv)
		\coloneqq \intrtM F_t(\boldv,\boldw) \,\textup{d}\boldw.
	\end{equation}
	Recall that the relative entropy $\Ent$ of a probability density function $f\in\LOneKac$ with respect to an equilibrium state $\gamma$ is defined by
	\begin{equation}
		\Ent(f|\gamma)
		\coloneqq \intrtN f(\boldv) \ln\left(\frac{f(\boldv)}{\gamma(\boldv)}\right) \,\textup{d}\boldv.
	\end{equation}
	Note that we use the inverted sign.
	
	In order to define the approach to equilibrium via the entropy of the Kac system, we need to specify the equilibrium state $\gamma$, that we compare the system to. If the Kac system is initially not in a Gaussian distributed state, we can not expect the system to reach a Gaussian distribution over time. However, in \cite{UniApproxMaxThermostatByHR} it is shown, that the Kac system coupled to a heat reservoir approximates the Kac system coupled to a thermostat uniformly in time in different norms for $\numhr\to\infty$. In other words, we can expect that the heat reservoir stays close to its initial state over time. Therefore, it is reasonable to use the initial state $\MwN$ of the heat reservoir as relative state for the entropy.
	
	As in the case of the thermostat, it is convenient to consider a ground state transformation (see also \cite[p. 3]{DecOfInfKac}). We define for $t \in [0,\infty)$ the function
	\begin{equation}
		\label{EqDefh_t}
		h_t(\boldv) \coloneqq \frac{f_t(\boldv)}{\MwN(\boldv)}.
	\end{equation}
	This allows us to write the entropy via
	\begin{equation}
		\label{EqEntht}
		\Ent\left( f_t | \MwN \right)
		= \intrtN h_t \ln(h_t) \,\MwN(\boldv) \textup{d}\boldv.
	\end{equation}
	Recall that the information of a probability density function $h$ with respect to the Gaussian measure $\MwN(\boldv)\textup{d}\boldv$ and $\sqrt{h}\in\HOneGauss$ is defined by
	\begin{equation}
		\InfTS(h)
		\coloneqq \intrtN \frac{|\nabla h(\boldv)|^2}{h(\boldv)} \, \MwN(\boldv)\textup{d}\boldv.
	\end{equation}
	Further, we can use the rotational invariance of the Gaussian $\MwN(\boldv)\MwN(\boldw)$ to get
	\begin{align*}
		h_t(\boldv)
		&\overset{\eqref{EqDefh_t}}{=} \frac{f_t(\boldv)}{\MwN(\boldv)}
		\overset{\eqref{EqDeff_t}}{=} \intrtM \frac{1}{\MwN(\boldv)} F_t(\boldv,\boldw) \,\textup{d}\boldw \\
		&\overset{\eqref{EqMasterEqHeatRes}}{=} \intrtM \frac{1}{\MwN(\boldv)} e^{\calL t}\left[f_0(\boldv)\MwN(\boldw)\right](\boldv,\boldw) \,\textup{d}\boldw \\
		&\overset{\eqref{EqDefh_t}}{=} \intrtM \frac{1}{\MwN(\boldv)} e^{\calL t}\left[h_0(\boldv)\MwN(\boldv)\MwN(\boldw)\right] \,\textup{d}\boldw \\
		&= \intrtM e^{\calL t}[h_0](\boldv,\boldw) \,\MwN(\boldw)\textup{d}\boldw \\
		&= \MargG e^{\calL t}[h_0](\boldv),
		\label{EqHtEvo} \numberthis
	\end{align*}
	where $\MargG$ is the marginal with respect to the Gaussian measure $\MwN(\boldw)\textup{d}\boldw$. Note that we considered the function $h_0:\bbR^{\di\numkac} \to \bbR$ as a function $\tilde{h}_0: \bbR^{\di(\numkac+\numhr)} \to \bbR$, where $\tilde{h}_0(\boldv,\boldw) \coloneqq h_0(\boldv)$.

	\subsection{Kinetic Energy}

	There is a one-to-one correspondence between the inverse temperature $\invtemp$ of a Gaussian distributed system and the kinetic energy
	\begin{equation}
	\label{EqEkinToInvTemp}
	\Ekin(\invtemp)
	\coloneqq \frac{1}{2} \intrtN \boldv^2 \left( \frac{\invtemp}{2\pi} \right)^\frac{\di\numkac}{2} e^{-\frac{\invtemp}{2} \boldv^2} \,\textup{d}\boldv
	= \frac{\di\numkac}{2\invtemp}.
	\end{equation}
	Thus, it is interesting to investigate the time evolution of the kinetic energy. We consider
	\begin{itemize}
		\item the total kinetic energy $\Ekin$ including the Kac system and the heat reservoir,
		\item the kinetic energy of the Kac system $\Ekinkac$, and
		\item the kinetic energy of the heat reservoir $\Ekinhr$.
	\end{itemize}
	These are defined by
	\begin{equation}
	\label{EqDefEkinHR}
	\begin{aligned}
	\Ekinkac(t)
	&\coloneqq \frac{1}{2} \sum_{i=1}^\numkac \intrtNM u_i^2 F_t(\boldu) \,\textup{d}\boldu
	= \frac{1}{2} \intrtN \boldv^2 f_t(\boldv) \,\textup{d}\boldv, \\
	\Ekinhr(t)
	&\coloneqq \frac{1}{2} \sum_{j=\numkac+1}^{\numkac+\numhr} \intrtNM u_i^2 F_t(\boldu) \,\textup{d}\boldu, \\
	\Ekin(t)
	&\coloneqq \frac{1}{2} \intrtNM \boldu^2 F_t(\boldu) \,\textup{d}\boldu
	= \Ekinkac(t) + \Ekinhr(t).
	\end{aligned}
	\end{equation}
	The total kinetic energy stays constant, since the collision mechanism conserves the kinetic energy. We will therefore write $\Ekin(t) = \Ekin$. The time evolution of the kinetic energy of the Kac system and heat reservoir can be computed by taking the derivative, deriving an ODE, and solving the latter. This is summarized in the following lemma.
	\begin{lem}
		\label{LemEkinHR}
		The total kinetic energy stays constant over time and is given by
		\begin{equation}
		\Ekin = \frac{\di\numhr}{2\invtemp} + \frac{1}{2} \intrtN \boldv^2 f_0(\boldv) \,\textup{d}\boldv
		\end{equation}
		The evolution of the kinetic energy of the Kac system and of the heat reservoir are given by
		\begin{align}
		\Ekinkac(t)
		&= \left( \Ekinkac(0) - \frac{\numkac}{\numkac + \numhr} \Ekin \right) e^{-\frac{\mu}{2\di\numhr}(\numkac + \numhr)t} + \frac{\numkac}{\numkac + \numhr} \Ekin, \\
		\Ekinhr(t)
		&= \left( \Ekinhr(0) - \frac{\numhr}{\numkac + \numhr} \Ekin \right) e^{-\frac{\mu}{2\di\numhr}(\numkac + \numhr)t} + \frac{\numhr}{\numkac + \numhr} \Ekin.
		\end{align}
	\end{lem}
	
	Lemma \ref{LemEkinHR} shows that the kinetic energy of the Kac system will tend to $\frac{\numkac}{\numkac + \numhr} \Ekin$. The decay rate is proportional to the number of particles in the Kac system, and thus much faster as what we will get for the entropy decay rate. Further, in the limit $t\to\infty$, all particles will have on average the same kinetic energy. Since the heat reservoir is assumed to be much larger than the Kac system, it is reasonable to use the inverse temperature $\invtemp$ of the initial state of the heat reservoir for the equilibrium state, that we compare the Kac system to.

	\subsection{Main results}

	Our first main result is that the information decays exponentially down to a small number.
	
	\begin{theo}
		\label{TheoDecOfInfHR}
		Let $(F_t)_{t\geq0}\in\LOneKNM$ be a solution of the Kac master equation \eqref{EqMasterEqHeatRes} with symmetry condition \eqref{EqPmeaSym} that satisfies the initial condition \eqref{EqInitalStateF_0} with $\sqrt{f_0} \in \HOneKac$. Assume further, that $\Ekin(0)<\infty$. Then, the information of $(h_t)_{t\geq0}$ defined by \eqref{EqDefh_t} decays as follows
		\begin{equation}
			\InfTS(h_t)
			\leq \left[ \frac{\numkac}{\numkac+\numhr} + \frac{\numhr}{\numkac+\numhr} e^{-\frac{\mu}{\di} \frac{\numkac+\numhr}{\numhr} t} \right] \InfTS(h_0).
		\end{equation}
		The proof is given in section \ref{ProofTheoDecOfInfHR}.
	\end{theo}

	In other words, if $\sqrt{\varphi_0} \in \HOneGauss$, we obtain under the Kac evolution
	\begin{equation}
		\InfTS\left(\MargG e^{\calL t} \varphi_0\right)
		\leq \left[ \frac{\numkac}{\numkac+\numhr} + \frac{\numhr}{\numkac+\numhr} e^{-\frac{\mu}{\di} \frac{\numkac+\numhr}{\numhr} t} \right] \InfTS(\varphi_0),
	\end{equation}
	where $\MargG$ denotes the marginal with respect to the Gaussian measure $\MwN(\boldw)\textup{d}\boldw$, i.e.\
	\begin{equation}
		\label{EqDefMargG}
		[\MargG \varphi](\boldv)
		\coloneqq \int_{\bbR^{\di\numhr}} \varphi(\boldv,\boldw) \,\MwN(\boldw)\textup{d}\boldw.
	\end{equation}
	In particular, since $\sqrt{P_s h_0} \in \HOneGauss$ if $\sqrt{h_0}\in\HOneGauss$, the estimate in theorem \ref{TheoDecOfInfHR} holds as well for $\varphi_t \coloneqq \MargG e^{\calL t} P_s h_0$. 
	We will show in the following that the decay rate of the information transfers to the entropy. The connection is established via Ornstein-Uhlenbeck semigroup by the following formula given in \cite[p.4, eq. (12)]{DecOfInfKac}
	\begin{equation}
		\label{EqConnEntInfHR}
		\Ent(f_t|\MwN)
		= \frac{1}{\invtemp} \int_0^\infty \InfTS(P_s\TS_t) \,\textup{d}s.
	\end{equation}
	This was similarly done in \cite[Lemma 3.1]{DecOfInfKac}. The crucial point now is to show that the time evolution operator $\calL$ commutes with the Ornstein-Uhlenbeck semigroup.
	\begin{lem}
		\label{LemLPsCommute}
		The time evolution operator $\calL$ commutes with the Ornstein-Uhlenbeck semigroup on $\LOneKNM$, i.e.\
		\begin{equation}
		[\calL, P_s] = \calL P_s - P_s \calL = 0, \quad \text{for all } s \geq 0.
		\end{equation}
		The proof is given in section \ref{ProofLemLPsCommute}.
	\end{lem}
	The transfer of the decay rate follows as given in the theorem below.
	\begin{theo}
		\label{TheoEntEstByInfEstHR}		
		Let $(F_t)_{t\geq0}\in\LOneKNM$ be a solution of the Kac master equation \eqref{EqMasterEqHeatRes} with symmetry condition \eqref{EqPmeaSym} that satisfies the initial condition \eqref{EqInitalStateF_0} with $\sqrt{f_0} \in \HOneKac$ and $\Ekin(0)<\infty$. Let further $(f_t)_{t>0}$ and $(h_t)_{t\geq0}$ be defined by \eqref{EqDeff_t} and \eqref{EqDefh_t}. If we have an inequality for the information of the form
		\begin{equation}
			\label{EqTheoEntDecByInfDecAssumpt}
			\InfTS(\MargG e^{\calL t} P_s h_0)
			\leq c(t) \InfTS(P_s h_0),
		\end{equation}
		then we get an inequality for the entropy by
		\begin{equation}
			\Ent(f_t|\MwN)
			\leq c(t) \Ent(f_0|\MwN).
		\end{equation}
		
		\begin{proof}
			We claim that $P_s$ commutes with the time evolution $\MargG e^{\calL t}$, i.e.\
			\begin{equation}
				\label{EqProofDecayInfClaim}
				P_s \MargG e^{\calL t} h_0 = \MargG e^{\calL t} P_s h_0.
			\end{equation}
			The theorem follows from this claim by
			\begin{align*}
				\Ent(f_t|\MwN)
				&
				\overset{\eqref{EqConnEntInfHR}}{=} \frac{1}{\invtemp} \int_0^\infty \textup{d}s\, \InfTS(P_sh_t)
				\overset{\eqref{EqHtEvo}}{=} \frac{1}{\invtemp} \int_0^\infty \textup{d}s\, \InfTS(P_s \MargG e^{\calL t} h_0)
				\overset{\eqref{EqProofDecayInfClaim}}{=} \frac{1}{\invtemp} \int_0^\infty \textup{d}s\, \InfTS(\MargG e^{\calL t} P_sh_0) \\
				&\overset{\eqref{EqTheoEntDecByInfDecAssumpt}}{\leq}
				\frac{c(t)}{\beta} \int_0^\infty \InfTS(P_sh_0) \,\textup{d}s
				= c(t) S(f_0|\MwN).
				\numberthis
			\end{align*}
			To show the claim \eqref{EqProofDecayInfClaim}, we will implicitly consider $h_0(\boldv,\boldw) \coloneqq h_0(\boldv)$ to be a function on $\bbR^{\di(\numkac+\numhr)}$. Let us denote the Ornstein-Uhlenbeck semigroup on $\bbR^{\di\numkac}$ by $\OUone$ and on $\bbR^{\di(\numkac+\numhr)}$ by $\OUtwo$. It immediately follows that $\OUtwo h_0(\boldv,\boldw) = \OUone h_0(\boldv)$. Further, we obtain by definition that
			\begin{equation}
				\MargG\OUtwo h_0(\boldv)
				= \int_{\bbR^{\di\numhr}} \int_{\bbR^{\di(\numkac+\numhr)}} h(e^{-s}(\boldv,\boldw) + \sqrt{1-e^{-2s}}(\boldx,\boldy)) e^{-\pi(\boldx^2+\boldy^2+\boldw^2)} \,\textup{d}\boldx\,\textup{d}\boldy\,\textup{d}\boldw.
			\end{equation}
			The rotation transformation $\boldp \coloneqq e^{-s}\boldw + \sqrt{1-e^{-2s}}\boldy, \boldq \coloneqq -\sqrt{1-e^{-2s}}\boldw + e^{-s}\boldy$ has Jacobian determinant 1 and fulfills $\boldp^2+\boldq^2 = \boldx^2+\boldy^2$. Hence, we obtain
			\begin{align*}
				\MargG\OUtwo h_0(\boldv)
				&= \int_{\bbR^{\di\numhr}} \int_{\bbR^{\di(\numkac+\numhr)}} h(e^{-s}\boldv + \sqrt{1-e^{-2s}}\boldx, \boldp) e^{-\pi(\boldx^2+\boldp^2+\boldq^2)} \,\textup{d}\boldx\,\textup{d}\boldp\,\textup{d}\boldq \\
				&= \OUone \MargG h_0(\boldv).
				\numberthis
			\end{align*}
			Using lemma \ref{LemLPsCommute}, this proves the claim \eqref{EqProofDecayInfClaim} by
			\begin{equation}
				\OUone \MargG e^{\calL t} h_0
				= \MargG \OUtwo e^{\calL t} h_0
				= \MargG e^{\calL t} \OUtwo h_0
				= \MargG e^{\calL t} \OUone h_0.
			\end{equation}
		\end{proof}
	\end{theo}

	Our next main result is now a direct consequence of the theorem \ref{TheoDecOfInfHR} applied to the function $P_s h_0$ and theorem \ref{TheoEntEstByInfEstHR}.
	
	\begin{theo}
		\label{TheoDecEntHR}
		Let $(F_t)_{t\geq0}\in\LOneKNM$ be a solution of the Kac master equation \eqref{EqMasterEqHeatRes} with the symmetry condition \eqref{EqPmeaSym} that satisfies the initial condition \eqref{EqInitalStateF_0} with $\sqrt{f_0} \in \HOneKac$ and $\Ekin(0)<\infty$. Let further $(f_t)_{t>0}$ be defined by \eqref{EqDeff_t}. The entropy of the Kac system decays for $t\geq0$ by
		\begin{equation}
		\Ent(f_t|\MwN)
		\leq \left[ \frac{\numkac}{\numkac+\numhr} + \frac{\numhr}{\numkac+\numhr} e^{-\frac{\mu}{d} \frac{\numkac+\numhr}{\numhr} t} \right] \Ent(f_0|\MwN).
		\end{equation}
	\end{theo}
	It is interesting to note that for the limit $\numhr\to\infty$, we get the same decay rates for the entropy and information as in the case of the thermostat.
	
	As pointed out in \cite[p. 7]{EntDecKac}, we can deduce the classic Kac model with $\numkac+\numhr$ particles defined by
	\begin{equation}
		\calL_\textup{cl}
		\coloneqq \frac{2}{\numkac+\numhr-1} \sum_{1\leq i < j \leq \numkac+\numhr} (R_{ij} - \bbone)
	\end{equation}
	from the heat reservoir model by setting the parameters
	\begin{equation}
		\lambda_S
		\coloneqq \frac{2(\numkac-1)}{\numkac + \numhr -1}, \quad
		\lambda_R
		\coloneqq \frac{2(\numhr-1)}{\numkac + \numhr -1}, \quad
		\mu
		\coloneqq \frac{2\numhr}{\numkac + \numhr -1}.
	\end{equation}
	This gives us the following corollary:
	
	\begin{cor}
		Let $F_0(\boldv,\boldw) = f_0(\boldv)\MwN(\boldw)$ and
		\begin{equation}
			f_t(\boldv)
			\coloneqq \intrtM \left[e^{\calL_\textup{cl} t} F_0\right] \textup{d}\boldw,
			\quad
			h_t(\boldv)
			\coloneqq \frac{f_t(\boldv)}{\MwN(\boldv)}.
		\end{equation}
		Assume that $\sqrt{f_0}\in\HOneKac$ and $\Ekin(0)<\infty$. The decay of information and entropy is then given by
		\begin{align}
			\InfTS(h_t)
			&\leq \left[ \frac{\numkac}{\numkac+\numhr} + \frac{\numhr}{\numkac+\numhr} e^{-\frac{2(\numkac+\numhr)}{\numkac+\numhr - 1} t} \right] \InfTS(h_0), \\
			\Ent\left( f_t | \MwN \right)
			&\leq \left[ \frac{\numkac}{\numkac+\numhr} + \frac{\numhr}{\numkac+\numhr} e^{-\frac{2(\numkac+\numhr)}{\numkac+\numhr - 1} t} \right] \Ent\left(f_0|\MwN\right).
		\end{align}
	\end{cor}

\section{Discussion and future work}
	\label{SecDiscussion}

	The decay rates for the entropy and information for the Kac system coupled to a thermostat or a heat reservoir obtained in the theorems \ref{TheoDecayInf}, \ref{TheoDecayEntropy}, \ref{TheoDecOfInfHR} and \ref{TheoDecEntHR} are consistent with the decay rates proven in \cite{KacThermostat}, \cite{EntDecKac} and \cite{DecOfInfKac}.	It is interesting to note that the decay rates are quite uniform in the sense that they are independent of the collisions within the Kac system, the probability measure used to sample the scattering angle, as well as the inverse temperature of the thermostat. They are also essentially independent of the size of the Kac system.
	
	While the information and entropy of the Kac system coupled to the thermostat tends to zero, the information and entropy of the Kac system coupled to a heat reservoir only tends to a small number. This is due to the fact, that the Kac system in the model with the heat reservoir will in general not tend towards a Gaussian distribution anymore. However, in the limit $\numhr\to\infty$ we recover decay rate of the Kac system coupled to a thermostat. Therefore, the estimate is expected to be optimal in some sense.
	
	In comparison to \cite{KacThermostat} and \cite{EntDecKac} the approach via the information used in this paper and \cite{DecOfInfKac} provides a more natural and elegant method with simpler proofs and weaker assumptions. It also allows to treat multidimensional systems as well as the spherical Kac model (see \cite{DecOfInfKac} for the latter).
	
	
	For future work, more intricate collision mechanisms could be considered. For example, the swapping map as given in \cite[Eq. (A.9)]{StrConMaxMod}
	\begin{equation}
	(v,w,\sigma)
	\mapsto \left( \frac{v+w}{2} + \frac{|v-w|}{2} \sigma, \frac{v+w}{2} - \frac{|v-w|}{2} \sigma, \frac{v-w}{|v-w|} \right)
	\end{equation}
	could be used instead of the reflection map. Since this map is firstly not linear, and secondly only bijective if the scattering angle is included as third argument, it us harder to handle. In particular, it is not immediately clear, how to prove that the time evolution commutes with the Ornstein-Uhlenbeck semigroup. However, we would still expect exponential decay rates with possibly different constants. Further, the case of hard sphere collisions or Maxwellian molecules could be discussed as suggested in \cite[p. 853]{EntDecKac}.
	
	In \cite[p. 19]{KacThermostat}, it was proposed to consider a Kac system coupled to two heat reservoirs at different temperatures. Using the information approach of this paper, it could be possible to obtain results about the decay of entropy for that model. The interesting part would be that the two big heat reservoirs would merge over time into one heat reservoir on a much larger time scale than it takes the smaller Kac system to tend towards the equilibrium state. However, the distribution of the merged reservoirs would not be a Gaussian distribution anymore. If a Gaussian state with averaged inverse temperature is used as relative state, we would expect to obtain an exponential decay for the entropy down to a small number.
	
	Arguably even more interesting is to consider the Kac system coupled to two thermostats at different temperatures. Even after long times, there would always be heat flowing into the Kac system from one thermostat and heat flowing from the Kac system to the other thermostat. Thus, we expect that the Kac system would tend towards a steady state instead of an equilibrium state. Therefore, the Kac system coupled to two thermostats might not be an approximation of the Kac system coupled to two heat reservoirs.
	
	Finally, it is open to our knowledge to rigorously derive the true distributions of the equilibrium or steady states, respectively. For example, we do not know the equilibrium state in the case of two coupled heat reservoirs with different temperatures under the Kac evolution, or the distribution of the steady state occuring in the case of a Kac system coupled to two thermostats.

	

\section{Proofs}

	\subsection{Proof of theorem \ref{TheoDecayInf}}
	\label{ProofTheoDecayInf}
	
	The proof consists of two parts.
	First, we estimate how pair collisions decrease the information of the Kac system. That is, we consider the collision of either two particles in the Kac system, or a particle in the Kac system and a particle in the thermostat.
	
	Second, we use the convexity of the information to break down the time evolution into single pair collisions. For this, we repeatedly write the solution $\TS_t$ as convex combination of simpler terms, and use Jensen's inequality.
	
	\begin{lem}
		\label{LemInfQ}
		The collisions of the particles in the Kac system with each other do not increase the information of the system. I.e.\ for all initial conditions $\sqrt{\TS_0}\in\HOneGauss$ the following estimate holds true:
		\begin{equation}
		\InfTS(Q\TS_0)
		\leq \InfTS(\TS_0).
		\end{equation}
		
		\begin{proof}
			Using the convexity of the information, we get
			\begin{align*}
			\InfTS(Q\TS_0)
			&= \InfTS\left( \binom{\numkac}{2} \sum_{i<j}^\numkac \intbbS \textup{d}\pmea(\sigma)\, \TS_0 \circ \ColMatij \right) \\
			&\leq \binom{\numkac}{2} \sum_{i<j}^\numkac \intbbS \textup{d}\pmea(\sigma)\, \InfTS\left( \TS_0 \circ \ColMatij \right).
			\numberthis
			\end{align*}
			Since the matrix $\ColMat = \begin{pmatrix} \bbone- \sigma\otimes\sigma & \sigma\otimes\sigma \\ \sigma\otimes\sigma & \bbone-\sigma\otimes\sigma \end{pmatrix}$ fulfills $\ColMat^2=\bbone$, the only possible eigenvalues of $\ColMat$ are $\pm 1$. This implies
			\begin{equation}
			\left|\left| \ColMatij \right|\right| = \max_{\lambda \in \sigma\left(\ColMatij\right)} |\lambda| = 1.
			\end{equation}
			This completes the proof by
			\begin{equation}
				\InfTS\left( \TS_0 \circ \ColMatij \right)
				= \int \MwN(\boldv)\textup{d}\boldv\, \frac{\left|\ColMatij \cdot \nabla \TS_0\left(\ColMatij \boldv\right)\right|^2}{\TS_0\left(\ColMatij \boldv\right)}
				= \InfTS(\TS_0).
				\label{EqInfColMatij}
			\end{equation}
		\end{proof}
	\end{lem}
	
	\begin{lem}
		\label{LemInfTj}
		The collision of a particle in the Kac system with the thermostat decreases its information by at least a factor of $\nicefrac{1}{d}$. The information of the other particles remains unchanged. I.e.\ for all initial conditions $\sqrt{\TS_0}\in\HOneGauss$, we have the estimate
		\begin{equation}
		\InfTS(T_j\TS_0)
		\leq \InfTS(\TS_0) - \frac{1}{\di} \intrtN \MwS(\boldv)\textup{d}\boldv\, \frac{\left|(\nabla\TS_0(\boldv))_j\right|^2}{\TS_0(\boldv)},
		\end{equation}
		where $(\nabla\TS_0(\boldv))_j\in\bbR^\di$ denotes the $j$-th three dimensional block component.
		
		\begin{nota*}
			To separate the action on the particles in the Kac system, we write
			\begin{equation}
			\label{EqColMatRel}
			\boldM_\sigma^{(j,\numkac+1)} (\boldv,w) = \left(\NColj \boldv + \wColj, \NCol w + \sigma\otimes\sigma[v_j]\right)
			\end{equation}
			where
			\begin{equation}
			\label{EqDefNColThermostat}
			\begin{split}
			\NCol
			&\coloneqq \bbone-\sigma\otimes\sigma \\
			\NColi
			&\coloneqq \diag(\bbone,\dots,\bbone,\underbrace{\bbone-\sigma\otimes\sigma}_{\text{i-th entry}},\bbone,\dots,\bbone), \\
			\boldw_\sigma^{(i)}
			&\coloneqq (0,\dots,0,\underbrace{\sigma\otimes\sigma[v_j]}_{\text{i-th entry}},0,\dots,0).
			\end{split}
			\end{equation}
			Note that $\bbone$ is the identity matrix in $\di$ dimensions and
			\begin{equation}
			{\NColi}^2 = \NColi
			\end{equation}
			This leads to the representation
			\begin{equation}
			R_j[f]
			= \intrt \textup{d}w\, \intbbS \textup{d}\pmea(\sigma)\, \MwS\left(\NCol w + \sigma\otimes\sigma[v_j] \right) f_t\left(\NColj \boldv + \wColj\right).
			\end{equation}
			For shorter notation, we write
			\begin{equation}
			\boldv^* \coloneqq \Proj \boldM_\sigma^{(j,\numkac+1)} \begin{pmatrix} \boldv \\ w \end{pmatrix} = \NColj \boldv + \wColj.
			\end{equation}
		\end{nota*}
		
		\begin{proof}
			Using the Cauchy-Schwarz inequality we get
			\begin{align*}
			\nabla[T_j\TS_0]
			&= \int \MwS(w) \textup{d}w \intbbS \textup{d}\pmea(\sigma)\,
			\NColj \nabla\TS_0\left(\boldv^*\right) \\
			|\nabla[T_j\TS_0]|^2
			&= \left(
			\int \MwS(w)\textup{d}w \intbbS \textup{d}\pmea(\sigma)\, \frac{\NColj \nabla\TS_0\left(\boldv^*\right)}{\sqrt{\TS_0\left(\boldv^*\right)}} \cdot \sqrt{\TS_0\left(\boldv^*\right)}
			\right)^2 \\
			&\leq \int \MwS(w) \textup{d}w \intbbS \textup{d}\pmea(\sigma)\, \frac{\left|\NColj \nabla\TS_0\left(\boldv^*\right)\right|^2}{\TS_0\left(\boldv^*\right)}
			\cdot T_j\TS_0.
			\numberthis
			\end{align*}
			Further since ${\NColj}^2 = \NColj$, we get
			\begin{align*}
			\left|\NColj \nabla\TS_0\right|^2
			&= \nabla \TS_0^T \NColj \nabla\TS_0
			= |\nabla\TS_0|^2 - |\sigma(\nabla\TS_0)_j|^2.
			\numberthis
			\end{align*}
			Therefore, we get with the transformation
			\begin{equation*}
			(\boldp,q) \coloneqq \boldM_\sigma^{(j,\numkac+1)} (\boldv,w)
			\end{equation*}
			that the information decreases by
			\begin{align*}
			\InfTS(T_j\TS_0)
			&\leq \intrtN \MwN(\boldp)\textup{d}\boldp \intrt \MwS(q)\textup{d}q \intbbS \textup{d}\pmea(\sigma)\,
			\frac{\left|\NColj \nabla\TS_0(\boldp)\right|^2}{\TS_0(\boldp)} \\
			&= \intrtN \MwN(\boldp)\textup{d}\boldp \intrt \MwS(q)\textup{d}q \intbbS \textup{d}\pmea(\sigma)\,
			\frac{|\nabla\TS_0|^2 - |\sigma(\nabla\TS_0)_j|^2}{\TS_0} \\
			&\leq \InfTS(\TS_0) - \frac{1}{\di} \intrtN \MwN(\boldp)\textup{d}\boldp\, \frac{|(\nabla\TS_0)_j|^2}{\TS_0}.
			\end{align*}
		\end{proof}
	\end{lem}
	
	\begin{rem}
		\label{RemInfT}
		It follows using the convexity of the information from the lemma \ref{LemInfTj} that
		\begin{equation}
		\InfTS\left( T \TS_0 \right)
		\leq \frac{1}{\numkac} \sum_{j=1}^\numkac \InfTS(T_j \TS_0)
		\leq \frac{1}{\numkac} \left(\numkac \InfTS(\TS_0) - \frac{1}{\di} \InfTS(\TS_0)\right) \\
		= \left( 1-\frac{1}{\di\numkac} \right) \InfTS(\TS_0).
		\numberthis
		\end{equation}
		In other words, the operator $T$ decreases the information by a factor of at least $\left(1-\frac{1}{\di\numkac}\right)$.
	\end{rem}
	
	Next, we can prove theorem \ref{TheoDecayInf} by repeatedly applying the convexity of the information to break down the information into simpler functions.
	
	First, we write the solution $\TS_t$ as a convex series
	\begin{align}
		\TS_t
		= e^{-(\lambda + \mu)\numkac t} \sum_{k=0}^\infty \frac{((\lambda + \mu)\numkac t)^k}{k!} \left( \frac{\lambda}{\lambda + \mu} Q + \frac{\mu}{\lambda + \mu} T \right)^k \TS_0.
	\end{align}
	Hence, we get for the information of $\TS_t$ that
	\begin{equation}
		\InfTS(\TS_t)
		\leq e^{-(\lambda + \mu)\numkac t} \sum_{k=0}^\infty \frac{((\lambda + \mu)t)^k}{k!} \InfTS\left(\left( \frac{\lambda}{\lambda + \mu} Q + \frac{\mu}{\lambda + \mu} T \right)^k \TS_0\right).
	\end{equation}
	Noting that $\frac{\lambda}{\lambda + \mu} Q + \frac{\mu}{\lambda + \mu} T$ is also a convex combination, we get using lemma \ref{LemInfQ} and remark \ref{RemInfT} that
	\begin{align*}
		\InfTS\left( \left(\frac{\lambda}{\lambda + \mu} Q + \frac{\mu}{\lambda + \mu} T\right) \TS_0 \right)
		&\leq \frac{\lambda}{\lambda + \mu} \InfTS(Q\TS_0) + \frac{\mu}{\lambda + \mu} \InfTS(T\TS_0) \\
		&\leq \left( 1- \frac{\mu}{\di\numkac(\lambda + \mu)} \right) \InfTS(\TS_0).
		\numberthis
	\end{align*}
	The claim now follows from the evaluation of the series
	\begin{align}
		\InfTS(\TS_t)
		\leq e^{-(\lambda + \mu)\numkac t} \sum_{k=0}^\infty \frac{((\lambda + \mu)\numkac t)^k}{k!} \left( 1-\frac{\mu}{\di\numkac(\lambda + \mu)} \right)^k \InfTS(\TS_0)
		= e^{-\frac{1}{\di} \mu t} \InfTS(\TS_0).
		\numberthis
	\end{align}

	\subsection{Proof of theorem \ref{TheoCommOUTimeEvolution}}
	\label{ProofTheoCommOUTimeEvolution}

	It is sufficient to prove, that the Ornstein-Uhlenbeck semigroup commutes with the operators $Q$ and $T_j$ for $j\in\{1,\dots,\numkac\}$.
	
	\subparagraph{Step 1:}
	
	We prove first, that the operator $Q$ commutes with the Ornstein-Uhlenbeck semigroup, i.e.\ for all $s\geq 0$ we have
	\begin{equation}
	QP_s = P_sQ.
	\end{equation}
	
	We write the transformation of the collisions using the matrix $\boldv_{i,j}(\sigma) = \ColMatij \boldv$ as described in \ref{EqDefColMatij}. The transformation $\boldy = \ColMatij \boldx$ yields
	\begin{align*}
	P_sQ[\TS]
	&= \binom{\numkac}{2}^{-1} \intrtN \MwN(\boldx)\textup{d}\boldx \sum_{i<j}^\numkac \intbbS \textup{d}\pmea(\sigma)\,
	\TS\Bigr( \underbrace{\ColMatij (e^{-s} \boldv + \sqrt{1-e^{-2s}} \boldx)}_{=e^{-s} \ColMatij \boldv + \sqrt{1-e^{-2s}} \ColMatij \boldx} \Bigr) \\
	&= \binom{\numkac}{2}^{-1} \intrtN \MwN(\boldy)\textup{d}\boldy \sum_{i<j}^\numkac \intbbS \textup{d}\pmea(\sigma)\, 
	\TS\left( e^{-s} \ColMatij \boldv + \sqrt{1-e^{-2s}} \boldy \right) \\
	&= QP_s[\TS].
	\numberthis
	\end{align*}
	
	\subparagraph{Step 2:}
	
	Next, we prove that the operators $T_j$ commute with the Ornstein-Uhlenbeck semigroup.
	
	We interpret any function $\TS^{(1)}\in L^1\left(\bbR^{\di\numkac},\MwN(\boldv)\textup{d}\boldv\right)$ as a function in two variables
	\begin{equation}
	\TS^{(2)}: \bbR^{\di\numkac}\times\bbR^\di\to\bbR, (\boldv,w) \mapsto \TS^{(1)}(\boldv).
	\end{equation}
	This allows us to write
	\begin{equation}
	T_j[ \TS^{(1)}](\boldv)
	= \intrt \MwS(w)\textup{d}w \intbbS \textup{d}\pmea(\sigma)\, \TS^{(2)}\left( \boldM_\sigma^{(j,\numkac+1)}(\boldv,w) \right).
	\end{equation}
	With the definitions
	\begin{align}
	\calU\big[\TS^{(2)}\big](\boldv,w)
	&\coloneqq \intbbS \textup{d}\pmea(\sigma)\, \TS^{(2)}\left( \boldM_\sigma^{(j,\numkac+1)}(\boldv,w) \right), \\
	\MargG \big[\TS^{(2)}\big](\boldv)
	&\coloneqq \intrt \MwS(w)\textup{d}w\, \TS^{(2)}(\boldv,w),
	\end{align}
	we get
	\begin{equation}
	T_j\big[\TS^{(1)}\big]
	= \MargG \calU\big[\TS^{(1)}\big].
	\end{equation}
	Similarly, we denote with $\OUone$ the $\di\numkac$-dimensional, and with $\OUtwo$ the $\di(\numkac+1)$-dimensional Ornstein-Uhlenbeck semigroup. Using $\OUalpha \coloneqq e^{-s}$ and $\OUbeta\coloneqq \sqrt{1-e^{-2s}}$ for shorter notation, we get
	\begin{align*}
	\OUtwo\TS^{(2)}(\boldv,w)
	&= \intrtN \MwS(\boldx)\textup{d}\boldx\intrt\MwS(y)\textup{d}y\, \TS^{(2)}(\alpha(\boldv,w) + \OUbeta(\boldx,y)) \\
	&= \intrtN \MwS(\boldx)\textup{d}\boldx\, \TS^{(1)}(\alpha \boldv + \OUbeta\boldx) \\
	&= \OUone \TS^{(1)}(\boldv).
	\numberthis
	\end{align*}
	The commutation relation between the Ornstein-Uhlenbeck semigroup and the marginal $\MargG$ is given by
	\begin{align*}
	\OUone\MargG[\TS^{(2)}](\boldv)
	&= \intrtN \MwS(\boldx)\textup{d}x \intrt \MwS(p)\textup{d}p\, \TS^{(2)}(\alpha \boldv + \OUbeta \boldx, p) \\
	&= \intrtN \MwS(\boldx)\textup{d}x \intrt \MwS(p)\textup{d}p \intrt \MwS(q)\textup{d}q\, \TS^{(2)}(\alpha \boldv + \OUbeta \boldx, p) \\
	&= \intrtN \MwS(\boldx)\textup{d}x \intrt \MwS(w)\textup{d}w \intrt \MwS(y)\textup{d}y\, \TS^{(2)}(\alpha \boldv + \OUbeta \boldx, \alpha w + \OUbeta y) \\
	&= \intrt \MwS(w)\textup{d}w\, \OUtwo[\TS^{(2)}](\boldv,w) \\
	&= \MargG\OUtwo[\TS^{(2)}](\boldv,w),
	\numberthis
	\end{align*}
	where we used the rotation transformation
	\begin{equation*}
	\begin{pmatrix} p \\ q \end{pmatrix} \coloneqq \begin{pmatrix} \alpha & \OUbeta \\ -\OUbeta & \alpha \end{pmatrix} \begin{pmatrix} w \\ y \end{pmatrix}.
	\end{equation*}
	Finally, the operator $U$ commutes with the $\di(\numkac+1)$-dimensional Ornstein-Uhlenbeck semigroup:
	\begin{align*}
	\calU\OUtwo&[\TS^{(2)}]
	= \intbbS \textup{d}\pmea(\sigma)\intrtN \MwS(\boldx)\textup{d}\boldx\intrt \MwS(y)\textup{d}y\,
	\TS^{(2)}\left( \boldM_\sigma^{(j,\numkac+1)}(\alpha(\boldv,w)+\OUbeta(\boldx,y)\right) \\
	&= \intbbS \textup{d}\pmea(\sigma)\intrtN \MwS(\boldx)\textup{d}\boldx\intrt \MwS(y)\textup{d}y\,
	\TS^{(2)}\left( \alpha\boldM_\sigma^{(j,\numkac+1)}(\boldv,w)+\OUbeta\boldM_\sigma^{(j,\numkac+1)}(\boldx,y) \right) \\
	&= \intbbS \textup{d}\pmea(\sigma)\intrtN \MwS(\boldx)\textup{d}\boldx\intrt \MwS(y)\textup{d}y\,
	\TS^{(2)}\left( \alpha\boldM_\sigma^{(j,\numkac+1)}(\boldv,w)+\OUbeta (\boldp,q) \right) \\
	&= \OUtwo\calU[\TS^{(2)}](\boldv,w),
	\numberthis				
	\end{align*}
	where we used the transformation
	\begin{equation*}
	\begin{pmatrix} \boldp \\ q \end{pmatrix} \coloneqq \boldM_\sigma^{(j,\numkac+1)} \begin{pmatrix} \boldx \\ y \end{pmatrix}.
	\end{equation*}
	The claim now follows by
	\begin{equation*}
	\OUone T_j [\TSone]
	= \OUone \MargG \calU [\TStwo]
	= \MargG \OUtwo \calU [\TStwo]
	= \MargG \calU \OUtwo [\TStwo]
	= T_j \OUone [\TSone].
	\end{equation*}

		\subsection{Proof of theorem \ref{TheoDecOfInfHR}}
		\label{ProofTheoDecOfInfHR}
		
		In a first remark, we break down the time evolution $e^{\calL t}$ into a convex series over collision histories, i.e.\ sequences of collisions. The proof then consists of two parts. First, we use the convexity of the information to break down the information decay of the time evolution into the information decay caused by collision histories. This reduces the problem to the derivation of a sum rule for some explicitly given matrix $K$, which we will derive in the second part.
		
		\begin{rem}[Representation as collision histories]
			\label{RemCollHistHR}
			We can represent the time evolution semigroup $\left(e^{\calL t}\right)_{t\geq0}$ in terms of collision histories, i.e.\ we can write $e^{\calL t}$ as convex series over products of pair collisions. For this, we write (see \cite[p. 7]{DecOfInfKac})
			\begin{equation}
			\calL = \Lambda(Q-\bbone)
			\end{equation}
			with
			\begin{align}
			\Lambda
			&\coloneqq \frac{\lambda_S \numkac}{2} + \frac{\lambda_R \numhr}{2} + \mu\numkac  \\
			Q
			&\coloneqq
			\frac{\lambda_S}{\Lambda(\numkac-1)} \sum_{1\leq i < j \leq \numkac} R_{ij}
			+ \frac{\lambda_R}{\Lambda(\numhr-1)} \sum_{\numkac<i<j\leq \numkac+\numhr} R_{ij}
			+ \frac{\mu}{\Lambda \numhr} \sum_{i=1}^\numkac \sum_{j=\numkac+1}^{\numkac+\numhr} R_{ij}.
			\end{align}
			This allows us to write $e^{\calL t}$ as the convex series
			\begin{equation}
			e^{\calL t}
			= e^{-\Lambda t} \sum_{k=0}^\infty \frac{(\Lambda t)^k}{k!} Q^k,
			\end{equation}
			Next, we want to express the powers $Q^k$ in terms of collision histories, i.e.\ by combinations of products of pair collisions. We introduce some notation: The index set $I$ denotes the set of collision pairs:
			\begin{equation}
			I \coloneqq \{(i,j): 1\leq i < j \leq \numkac + \numhr \}.
			\end{equation}
			For $\alpha \in I$ we define the coefficients
			\begin{equation}
			\lambda_\alpha
			\coloneqq \begin{cases}
			\frac{\lambda_S}{\Lambda(\numkac-1)}, &1 \leq i < j \leq \numkac, \\
			\frac{\lambda_R}{\Lambda(\numhr-1)}, &\numkac < i < j \leq \numkac+\numhr, \\
			\frac{\mu}{\Lambda \numhr}, &1 \leq i \leq \numkac < j \leq \numkac+\numhr.
			\end{cases}
			\end{equation}
			For $\balpha= (\alpha_1,\dots,\alpha_k)\in I^k$ we set $\lambda^{\balpha} \coloneqq \lambda_{\alpha_k} \cdots \lambda_{\alpha_1}$ and $R^{\balpha} \coloneqq R_{\alpha_k} \cdots R_{\alpha_1}$. Now, we can write the powers $Q^k$ by
			\begin{align*}
			Q^k
			&= \sum_{\balpha\in I^k} \lambda^{\balpha} R^{\balpha}.
			\numberthis
			\end{align*}
			Note that $\sum_{\alpha\in I} \lambda_\alpha = 1 = \sum_{\balpha \in I^k} \lambda^{\balpha}$ is again a convex combination. Therefore, the semigroup is given by the following convex series of pair collision operators:
			\begin{equation}
			\label{EqColHis}
			e^{\calL t}
			= e^{-\Lambda t} \sum_{k=0}^\infty \frac{(\Lambda t)^k}{k!} \sum_{\balpha \in I^k} \lambda^{\balpha} R^{\balpha}.
			\end{equation}
		\end{rem}
	
		\begin{lem}
			The information decay is given by
			\begin{equation}
				\label{EqProofInfWithKHR}
				\InfTS(\TS_t|\MwN)
				\leq \intrtN \MwN(\boldv) \textup{d}\boldv\,
				\frac{\nabla h_0(\boldv)^T \boldK \nabla h_0(\boldv)}{h_0(\boldv)}
			\end{equation}
			where
			\begin{equation}
				\label{EqHRK}
				\boldK
				\coloneqq e^{-\lambda t} \sum_{k=0}^\infty \frac{(\Lambda t)^k}{k!}
				\sum_{\balpha \in I^k} \lambda^{\balpha}
				\intkbbS \textup{d}\pmea(\bsigma) \,
				\boldA_k^T(\balpha,\bsigma) \boldA_k(\balpha,\bsigma).
			\end{equation}
			
			\begin{proof}
				Inserting the representation formula \eqref{EqColHis} in equation \ref{EqHtEvo}, we get
				\begin{align*}
				h_t(\boldv)
				&= \MargG e^{\calL t} [h_0\circ \Proj]
				= e^{-\Lambda t} \sum_{k=0}^\infty \frac{(\Lambda t)^k}{k!} \sum_{\balpha \in I^k} \lambda^{\balpha}
				\MargG R^{\balpha}[h_0\circ\Proj](\boldv) \\
				&= e^{-\Lambda t} \sum_{k=0}^\infty \frac{(\Lambda t)^k}{k!} \sum_{\balpha \in I^k} \lambda^{\balpha} \intkbbS \textup{d}\rho(\bsigma)\,
				\underbrace{\MargG [h_0\circ\Proj]\left(\boldM_{\sigma_k}^{\alpha_k} \cdots \boldM_{\sigma_1}^{\alpha_1} (\boldv,\boldw)^T \right)}_{\eqqcolon h_{k,\balpha,\bsigma}(\boldv)},
				\numberthis
				\end{align*}
				where $\pmea(\bsigma)=(\pmea(\sigma_1),\dots,\pmea(\sigma_k))$.	The convexity of the information implies by Jensen's inequality that
				\begin{equation}
					\label{EqProofInfDecHR}
					\calI(h_t|\MwN)
					\leq e^{-\Lambda t} \sum_{k=0}^\infty \frac{(\Lambda t)^k}{k!} \sum_{\balpha\in I^k} \lambda^{\balpha} \intkbbS \textup{d}\pmea(\bsigma)\,
					\calI\left( h_{k,\balpha,\bsigma} | \MwN \right).
				\end{equation}
				Next, we separate the action of the rotations by denoting
				\begin{equation}
				\begin{pmatrix}
				\boldA_k(\balpha,\bsigma) & \boldB_k(\balpha,\bsigma) \\
				\boldC_k(\balpha,\bsigma) & \boldD_k(\balpha,\bsigma)
				\end{pmatrix}
				\coloneqq \boldM_{\sigma_k}^{(\alpha_k)} \cdots \boldM_{\sigma_1}^{(\alpha_1)},
				\end{equation}
				where $\boldA_k(\balpha,\bsigma) \in \bbR^{\di\numkac\times\di\numkac}, \boldB_k(\balpha,\bsigma) \in \bbR^{\di\numkac\times\di\numhr}, \boldC_k(\balpha,\bsigma) \in \bbR^{\di\numhr\times\di\numkac}, \boldD_k(\balpha,\bsigma) \in \bbR^{\di\numhr\times\di\numhr}$. This allows us to observe, that
				\begin{align*}
				h_{k,\balpha,\bsigma}(\boldv)
				&= \intrtM \MwN(\boldw)\textup{d}\boldw\, h_0\left( \begin{pmatrix} \bbone & 0 \end{pmatrix} 
				\begin{pmatrix}
				\boldA_k & \boldB_k \\
				\boldC_k & \boldD_k
				\end{pmatrix}
				\begin{pmatrix} \boldv \\ \boldw \end{pmatrix}\right) \\
				&= \intrtM \MwN(\boldw)\textup{d}\boldw\, h_0\left( \boldA_k \boldv + \boldB_k \boldw \right)
				\numberthis
				\end{align*}
				is independent of $\boldC_k$ and $\boldD_k$. Further, this representation of $h_{k,\balpha,\bsigma}$ allows an easy computation of its gradient. Using Jensen's inequality in the first step, the information of $h_{k,\balpha,\bsigma}$ is given by
				\begin{align*}
				\InfTS(h_{k,\balpha,\bsigma}))
				&\leq \intrtNM 
				\frac{|\nabla_{\boldv} [h_0(\boldA_k \boldv + \boldB_k \boldw)]|^2}{h_0(\boldA_k \boldv + \boldB_k \boldw)} \,\MwN(\boldv)\MwN(\boldw)\, \textup{d}\boldv\,\textup{d}\boldw\\
				&= \intrtNM 
				\frac{|\boldA_k [\nabla h_0](\boldA_k \boldv + \boldB_k \boldw)]|^2}{h_0(\boldA_k \boldv + \boldB_k \boldw)} \,\MwN(\boldv)\MwN(\boldw) \,\textup{d}\boldv\,\textup{d}\boldw \\
				&= \intrtNM \frac{\nabla h_0(\boldp)^T \boldA_k^T \boldA_k \nabla h_0(\boldp)}{h_0(\boldp)} \,\MwN(\boldp)\MwN(\boldq)\, \textup{d}\boldp\,\textup{d}\boldq \\
				&= \intrtN \frac{\nabla h_0(\boldp)^T \boldA_k^T \boldA_k \nabla h_0(\boldp)}{h_0(\boldp)} \,\MwN(\boldp)\textup{d}\boldp,
				\numberthis
				\end{align*}
				where we used the transformation
				\begin{equation}
				\begin{pmatrix} \boldp \\ \boldq \end{pmatrix} \coloneqq \begin{pmatrix} \boldA_k & \boldB_k \\ \boldC_k & \boldD_k \end{pmatrix} \begin{pmatrix} \boldv \\ \boldw \end{pmatrix}.
				\end{equation}
			Inserting this formula in the inequality \eqref{EqProofInfDecHR} yields the claim.
			\end{proof}
		\end{lem}
		
		Computing the matrix $\boldK$ in the following lemma completes the proof.
		
		\begin{lem}
			\label{LemSumRuleK}
			The matrix $\boldK$ defined by \eqref{EqHRK} is given by
			\begin{equation}
			\boldK
			= \left[ \frac{\numkac}{\numkac+\numhr} + \frac{\numhr}{\numkac+\numhr} e^{-\frac{\mu(\numkac+\numhr)}{\di\numhr}t} \right] \bbone_{\di\numkac}.
			\end{equation}
			
			\begin{proof}
				First, we observe that $\boldA_k^T \boldA_k$ is the top left entry of the matrix
				\begin{equation}
				\begin{pmatrix} \boldA_k & \boldB_k \\ \boldC_k & \boldD_k \end{pmatrix}^T
				\begin{pmatrix} \bbone & 0 \\ 0 & 0 \end{pmatrix}
				\begin{pmatrix} \boldA_k & \boldB_k \\ \boldC_k & \boldD_k \end{pmatrix}.
				\end{equation}
				That is, we can write
				\begin{equation}
				\boldA_k^T \boldA_k
				= \Proj \boldM_{\sigma_1}^{(\alpha_1)}\dots \boldM_{\sigma_k}^{(\alpha_k)}
				\begin{pmatrix} \bbone & 0 \\ 0 & 0 \end{pmatrix}
				\boldM_{\sigma_k}^{(\alpha_k)}\dots \boldM_{\sigma_1}^{(\alpha_1)} \Proj^T.
				\end{equation}
				Next, we will iteratively integrate out the factors $\boldM_{\sigma_k}^{(\alpha_k)}
				\begin{pmatrix} \bbone & 0 \\ 0 & 0 \end{pmatrix}
				\boldM_{\sigma_k}^{(\alpha_k)}$ from the inside to the outside. For this, we note that pointwise integration in the matrix yields
				\begin{align*}
				\label{EqProofKCoeffTilde}
				\intbbS \textup{d}\pmea(\sigma)\,
				&M_\sigma \begin{pmatrix} m_1 \bbone_\di & 0 \\ 0 & m_2 \bbone_\di \end{pmatrix} M_\sigma \\
				&= \intbbS \textup{d}\pmea(\sigma)\,
				\begin{pmatrix} m_1\bbone_\di - (m_1-m_2)(\sigma\otimes\sigma) & 0 \\ 0 & m_2 \bbone_\di - (m_2-m_1)(\sigma\otimes\sigma) \end{pmatrix} \\
				&= \begin{pmatrix} \left(m_1 - \frac{1}{\di}(m_1-m_2)\right)\bbone_\di & 0 \\ 0 & \left(m_2 - \frac{1}{\di} (m_2-m_1)\right) \bbone_\di \end{pmatrix} \\
				&= \begin{pmatrix} \tilde{m_1}\bbone_\di & 0 \\ 0 & \tilde{m_2} \bbone_\di \end{pmatrix},
				\numberthis
				\end{align*}
				with
				\begin{equation}
				\tilde{m}_1 \coloneqq m_1 - \frac{1}{d}(m_1-m_2),\quad
				\tilde{m}_2 \coloneqq m_2 - \frac{1}{d}(m_2-m_1).
				\end{equation}
				This proves that the matrix always stays diagonal throughout the process and shows how the coefficients develop. In the following, we compute the sum over all collision pairs and integrals over all scattering angles for this expression with the general collision matrix $\boldM_\sigma^{(i,j)}$. We use the notation
				\begin{equation}
				\boldL(m_1,m_2)
				\coloneqq \begin{pmatrix} m_1\bbone_{\di\numkac} & 0 \\ 0 & m_2 \bbone_{\di\numhr} \end{pmatrix}.
				\end{equation}
				Further, we distinguish between the three cases of collisions in the Kac system, in the heat reservoir, and the interaction between the systems:
				\begin{equation}
				\sum_{\alpha\in I} \lambda_\alpha
				= \sum_{1\leq i < j \leq \numkac} \frac{\lambda_S}{\Lambda(\numkac-1)}
				+ \sum_{\numkac < i < j \leq \numkac+\numhr} \frac{\lambda_R}{\Lambda(\numhr-1)}
				+ \sum_{i=1}^\numkac \sum_{j=\numkac+1}^{\numkac+\numhr} \frac{\mu}{\Lambda \numhr}.
				\end{equation}
				
				\subparagraph{Step 1:}
				
				The collisions between particles in the Kac system give us
				\begin{align*}
				\frac{\lambda_S}{\Lambda(\numkac-1)} &\sum_{1\leq i < j \leq \numkac} \intbbS \textup{d}\pmea(\sigma)\, \ColMatij \boldL(m_1,m_2) \ColMatij \\
				&= \frac{\lambda_S}{\Lambda(\numkac-1)} \sum_{1\leq i < j \leq \numkac} \intbbS \textup{d}\pmea(\sigma) \begin{pmatrix} m_1 {\ColMatij}^2 & 0 \\ 0 & m_2 \bbone \end{pmatrix} \\
				&= \frac{\lambda_S \numkac}{2\Lambda} \boldL(m_1,m_2).
				\numberthis
				\end{align*}
				
				\subparagraph{Step 2:}
				
				The collisions between particles in the heat reservoir act similarly on $\boldL(m_1,m_2)$:
				\begin{align*}
				\frac{\lambda_R}{\Lambda(\numhr-1)} &\sum_{\numkac < i < j \leq \numkac+\numhr} \intbbS \textup{d}\pmea(\sigma) \ColMatij \boldL(m_1,m_2) \ColMatij \\
				&= \frac{\lambda_R}{\Lambda(\numhr-1)} \sum_{\numkac < i < j \leq \numkac+\numhr} \intbbS \textup{d}\pmea(\sigma) \begin{pmatrix} m_1 \bbone & 0 \\ 0 & m_2 {\ColMatij}^2 \end{pmatrix} \\
				&= \frac{\lambda_R \numhr}{2\Lambda} \boldL(m_1,m_2).
				\numberthis
				\end{align*}
				
				\subparagraph{Step 3:}
				
				Using \eqref{EqProofKCoeffTilde} we can compute the interaction between the two systems by
				\begin{align*}
				&\frac{\mu}{\Lambda \numhr} \sum_{i=1}^\numkac \sum_{j=\numkac+1}^{\numkac+\numhr} \intbbS \textup{d}\pmea(\sigma)\,
				\ColMatij \boldL(m_1,m_2) \ColMatij \\
				&= \frac{\mu}{\Lambda \numhr} \sum_{i=1}^\numkac
				\sum_{j=\numkac+1}^{\numkac+\numhr}
				\diag(m_1\bbone_\di,\dots,\underbrace{\tilde{m_1}\bbone_\di}_{i-\text{th entry}},\dots,m_1\bbone_\di,m_2\bbone_\di,\dots,\underbrace{\tilde{m_2}\bbone_\di}_{j-\text{th entry}},\dots,m_2\bbone_\di) \\
				&= \frac{\mu}{\Lambda \numhr} \begin{pmatrix}
				\left(\sum_{j=\numkac+1}^{\numkac+\numhr}\, (\numkac-1)m_1+\tilde{m}_1\right) \bbone_{\di\numkac} & 0 \\ 0
				&\left(\sum_{i=1}^\numkac\, (\numhr-1)m_2 + \tilde{m}_2\right) \bbone_{\di\numhr}
				\end{pmatrix} \\
				&= \frac{\mu}{\Lambda\numhr}
				\begin{pmatrix} \left(\numkac\numhr m_1 + \frac{\numhr}{\di}(m_2-m_1)\right)\bbone_{\di\numkac}
				& 0 \\ 0 &
				\left( \numkac\numhr m_2 + \frac{\numkac}{\di} (m_1-m_2\right)\bbone_{\di\numhr}
				\end{pmatrix} \\
				&= \frac{\mu\numkac}{\Lambda} \boldL(m_1,m_2) - \frac{\mu}{\di\Lambda\numhr} \boldL\big(\numhr(m_1-m_2),\numkac(m_2-m_1)\big).
				\numberthis
				\end{align*}
				
				From Step 1 to 3, we get the sum of the three terms to
				\begin{align*}
				\sum_{\alpha\in I} \lambda_\alpha \intbbS \textup{d}\pmea(\sigma) &\boldM_\sigma^{(\alpha)} \boldL(m_1,m_2) \boldM_\sigma^{(\alpha)} \\
				&= \frac{\lambda_S \numkac}{2\Lambda} \boldL(m_1,m_2)
				+ \frac{\lambda_R \numhr}{2\Lambda} \boldL(m_1,m_2)
				+ \frac{\mu\numkac}{\Lambda} \boldL(m_1,m_2) \\
				&\,- \frac{\mu}{\di\Lambda\numhr} \boldL\big(\numhr(m_1-m_2),\numkac(m_2-m_1)\big) \\
				&= \boldL(m_1,m_2) - \frac{\mu}{\di\Lambda\numhr} \boldL(\numhr(m_1-m_2),\numkac(m_2-m_1)) \\
				&= \boldL(m_1',m_2')
				\numberthis
				\end{align*}
				with
				\begin{equation}
				\begin{pmatrix} m_1' \\ m_2' \end{pmatrix}
				= \underbrace{\left( \bbone_2
					- \frac{\mu}{\di\Lambda\numhr} \begin{pmatrix} \numhr & - \numhr \\ -\numkac & \numkac \end{pmatrix} \right)}_{\eqqcolon \calP}
				\begin{pmatrix} m_1 \\ m_2 \end{pmatrix}
				= \calP \begin{pmatrix} m_1 \\ m_2 \end{pmatrix}.
				\end{equation}
				If we start with $m_1= 1, m_2 = 0$, and apply $\calP$ iteratively $k$-times, we get
				\begin{equation}
				\sum_{\balpha \in I^k} \lambda^{\balpha}
				\intkbbS \textup{d}\pmea(\bsigma) \,
				\boldA_k^T(\balpha,\bsigma) \boldA_k(\balpha,\bsigma)
				= e^{-\lambda t} \sum_{k=0}^\infty \frac{(\Lambda t)^k}{k!}
				\Proj \boldL\left(\calP^k\begin{pmatrix} 1 \\ 0 \end{pmatrix} \right) \Proj^T.
				\end{equation}
				To compute this expression, we decompose the start vector into the eigenvectors of $\calP$. The eigenvalues of $\calP$ are given by $\lambda_1 = 1, \lambda_2 = 1-\frac{\mu(\numkac+\numhr)}{\di\Lambda \numhr}$ with corresponding eigenvectors $v_1 = (1,1)^T, v_2 = \frac{1}{\numkac+\numhr} (\numhr,-\numkac)^T$. Using $(1,0)^T = \frac{\numkac}{\numkac+\numhr} v_1 + v_2$, we get
				\begin{equation}
				\calP^k\begin{pmatrix} 1 \\ 0 \end{pmatrix}
				= \frac{\numkac}{\numkac+\numhr} \begin{pmatrix} 1 \\ 1 \end{pmatrix} + \left( 1 - \frac{\mu(\numkac+\numhr)}{\di\Lambda \numhr} \right)^k \frac{1}{\numkac+\numhr} \begin{pmatrix} \numhr \\ - \numkac \end{pmatrix}.
				\end{equation}
				Taking the first entry, we get
				\begin{equation}
				\Proj \boldL\left(\calP^k\begin{pmatrix} 1 \\ 0 \end{pmatrix} \right) \Proj^T
				= \left( \frac{\numkac}{\numkac+\numhr} + \frac{\numhr}{\numkac+\numhr} \left(1-\frac{\mu(\numkac+\numhr)}{\di\Lambda \numhr} \right)^k\right) \bbone_{\di\numkac}.
				\end{equation}
				This proves the claim as follows:
				\begin{align*}
				\boldK
				&= e^{-\lambda t} \sum_{k=0}^\infty \frac{(\Lambda t)^k}{k!}
				\left( \frac{\numkac}{\numkac+\numhr} + \frac{\numhr}{\numkac+\numhr} \left(1-\frac{\mu(\numkac+\numhr)}{\di\Lambda \numhr} \right)^k\right) \bbone_{\di\numkac} \\
				&= \left[ \frac{\numkac}{\numkac+\numhr} + \frac{\numhr}{\numkac+\numhr} e^{-\frac{\mu(\numkac+\numhr)}{\di\numhr}t} \right] \bbone_{\di\numkac}.
				\end{align*}
			\end{proof}
		\end{lem}

		\subsection{Proof of lemma \ref{LemLPsCommute}}
			\label{ProofLemLPsCommute}
		
		Recall that for any $F\in L^1\left(\bbR^{\di(\numkac+\numhr)}, \MwN(\boldu) \textup{d}\boldu\right)$, the Ornstein-Uhlenbeck semigroup is defined by
		\begin{equation}
		P_s[F](\boldu)
		= \intrtNM F\left( e^{-s} \boldu + \sqrt{1-e^{-2s}} \boldw \right) \,\MwN(\boldx) \textup{d}\boldx.
		\end{equation}
		Using the change of variables $\boldy \coloneqq \ColMatij \boldx$, the rotational invariance of $\MwN$ and the linearity of the reflection map we get
		\begin{align*}
		P_sR_{ij}[F](\boldu)
		&= \intrtNM \MwN(\boldx) \textup{d}\boldx \intbbS \textup{d}\pmea(\sigma) F\left( \ColMatij \left(e^{-s} \boldu + \sqrt{1-e^{-2s}} \boldx \right) \right) \\
		&= \intrtNM \MwN(\boldx) \textup{d}\boldx \intbbS \textup{d}\pmea(\sigma) F\left( e^{-s} \ColMatij \boldu + \sqrt{1-e^{-2s}} \ColMatij \boldx \right) \\
		&= \intrtNM \MwN(\boldy) \textup{d}\boldy \intbbS \textup{d}\pmea(\sigma) F\left( e^{-s} \ColMatij \boldu + \sqrt{1-e^{-2s}} \boldy \right) \\
		&= R_{ij}P_s[F](\boldu),
		\numberthis
		\end{align*}
		for all $i,j \in \{1,\dots,\numkac+\numhr\}, i \neq j$. Since $\calL$ is a linear combination of operators $R_{ij}$, this proves the claim.
	
	\vspace{12pt}
	\appendix
	\addcontentsline{toc}{section}{Appendix}
	
	\Huge \textbf{Appendix} \normalsize

	\section{Technical remarks}
	
	In this chapter, we make some remarks on elementary facts that are either interesting by themselves or used in the main part with reference to this appendix for improved clarity.
	
	\begin{rem}[Collision process]
		\label{RemKacThermConservation} \hfill
		
		The collision transformation $(v_i,v_j) \mapsto \ColMat (v_i,v_j)$ of two particles within the Kac system is a momentum and kinetic energy conserving rotation. It follows by straight calculation that
		\begin{equation}
		\ColMat = \ColMat^T = \ColMat^{-1},
		\end{equation}
		which implies that
		\begin{equation}
		\det\left(\ColMat\right) = 1, \\
		v_i + v_j = v_i^* + v_j^*,
		v_i^2 + v_j^2 = \left(v_i^*\right)^2 + \left(v_j^*\right)^2,
		\end{equation}
		where $(v_i^*,v_j^*)^T = \ColMat (v_i,v_j)^T$ are the post-collisional velocity vectors. It follows, that
		\begin{equation}
		\label{EqColMech}
		\begin{aligned}
		\ColMatij = {\ColMatij}^T &= {\ColMatij}^{-1}, \quad
		&\det\left(\ColMatij\right) &= 1, \\
		\sum_{i=1}^\numkac \left(\ColMatij \boldv\right)_i &= \sum_{i=1}^\numkac \boldv_i, \quad
		&\left(\ColMatij \boldv\right)^2 &= \boldv^2.
		\end{aligned}
		\end{equation}
	\end{rem}

		\begin{rem}[Information of the transformed state]
		\label{RemInfRelTS} \hfill
		
		For both models, we can relate the information $\InfK(f_0)$ to the information of the transformed state $\InfTS(\TS_0)$ as follows. If the initial kinetic energy $\Ekin(0)<\infty$ is finite, we have
		\begin{equation}
		\sqrt{f_0}\in\HOneKac
		\quad \Leftrightarrow \quad
		\sqrt{\TS_0}\in\HOneGauss.
		\end{equation}
		If one of the statements is true, we get in particular that
		\begin{equation}
		\InfTS(\TS_t) = \InfK(f_t) + 2\invtemp^2 \left( \Ekin(t) - \frac{\di\numkac}{\invtemp} \right)
		\end{equation}
		Note that for the Kac model coupled to a thermostat, we get
		\begin{equation}
		\lim_{t\to\infty} \InfTS(h_t) = \lim_{t\to\infty}\InfK(f_t)
		\end{equation}
		since $\Ekin(t) \to \frac{\di\numkac}{\beta}$ by lemma \ref{LemEkinMom}.
		
		\begin{proof}
			It is clear that $f_0\in\LOneKac$ if and only if $\TS_0\in\LOneGauss$. By the definition, it is also clear that $\nabla f_0$ exists in a weak sense if and only if $\nabla \TS_0$ exists in a weak sense. Therefore, the following computation completes the proof:
			\begin{align*}
			\InfTS(\TS_t)
			&= \intrtN \frac{|\nabla \TS_t|^2}{\TS_t} \,\MwN(\boldv)\textup{d}\boldv
			= \intrtN \left| \frac{\nabla f_t \cdot \MwN + \invtemp v f_t \MwN}{\MwN^2} \right|^2 \frac{\MwN^2}{f_t} \,\textup{d}\boldv \\
			&= \intrtN \frac{|\nabla f_t + \invtemp \boldv f_t|^2}{f_t} \,\textup{d}\boldv \\
			&= \intrtN \frac{|\nabla f_t|^2 + 2\invtemp f_t \boldv \nabla f_t + \invtemp^2 \boldv^2 f_t^2}{f_t} \,\textup{d}\boldv \\
			&= \InfK(f_t) + 2\invtemp \intrtN \boldv \nabla f_t \,\textup{d}\boldv + \invtemp^2 \intrtN \boldv^2 f_t \,\textup{d}\boldv \\
			&= \InfK(f_t) - 2\invtemp \intrtN \underbrace{\nabla \cdot \boldv}_{=\di\numkac} f_t \,\textup{d}\boldv + 2\invtemp^2 \Ekin(t) \\
			&= \InfK(f_t) + 2 \invtemp^2 \left( \Ekin(t) - \frac{\di\numkac}{\invtemp} \right).
			\label{EqInfTransfThermo} \numberthis
			\end{align*}
		\end{proof}
	\end{rem}

	\begin{rem}[Regularity of the Kac system coupled to a thermostat]
		\label{RemKacThermoReg} \hfill
		
		Considering the Kac system coupled to a thermostat, we show that the operators $Q$ and $R$ are bounded with $||Q||=1, ||R||=1$. This implies that $\calL$ is bounded with $||\calL||
		\leq 2(\lambda + \mu) \numkac$. Therefore, the time evolution operator $\calL$ is by corollary 1.5 in \cite{EngelNagel} the generator of the exponential semigroup $\left(e^{\calL t}\right)_{t\geq0}$. In fact, we show that $\calL$ generates a contraction semigroup. Further, we show that $f_0$ is a probability density function if and only if $f_t$ is a probabilty density function for all $t\geq0$. This is due to the fact that the operators $Q$ and $R$ are averages over rotations.
		
		\begin{enumerate}[label=\roman*)]
			\item \label{RemKacThermoRegParti}
			The operator $Q$ is well-defined with $||Q|| \leq 1$, since for any $f\in\LOneKac$, we get with the transformation $\boldp\coloneqq\ColMatij \boldv$ that
			\begin{align*}
			||Qf||_1
			&\leq \binom{\numkac}{2}^{-1} \sum_{1\leq i < j \leq \numkac} \intrtN \textup{d}\boldv \intbbS \textup{d}\pmea(\sigma)\, \left| f\left( \ColMatij \boldv \right) \right| \\
			&= \binom{\numkac}{2}^{-1} \sum_{1\leq i < j \leq \numkac} \intrtN \textup{d}\boldp \intbbS \textup{d}\pmea(\sigma)\, \left| f\left( \boldp \right) \right| \\
			&= ||f||_1.
			\numberthis
			\end{align*}
			
			\item \label{RemKacThermoRegPartii}
			The operator $R$ is well-defined with $||R|| \leq 1$, since for any $f\in\LOneKac$, we get with the transformation
			\begin{equation*}
			\begin{pmatrix} \boldp \\ q \end{pmatrix} \coloneqq \boldM_\sigma^{(j,\numkac+1)} \begin{pmatrix} \boldv \\ w \end{pmatrix}
			\end{equation*}
			that
			\begin{align*}
			&||R_j f||_1 \\
			&\leq \intrtN \textup{d}\boldv \intrt \textup{d}w \intbbS \textup{d}\pmea(\sigma)\,
			\MwS\left( \left[ \boldM_\sigma^{(j,\numkac+1)} \begin{pmatrix} \boldv \\ w \end{pmatrix} \right]_{\numkac+1} \right)
			\left| f\left( \Proj \boldM_\sigma^{(j,\numkac+1)} \begin{pmatrix} \boldv \\ w \end{pmatrix} \right) \right| \\
			&= \intrtN \textup{d}\boldv \intrt \textup{d}w \intbbS \textup{d}\pmea(\sigma)\,
			\MwS(q) |f(\boldp)| \\
			&= ||f||_1.
			\numberthis
			\end{align*}
			
			\item \label{RemKacThermoRegPartiiiContSG}
			Next, we show that $\calL$ generates a contraction semigroup. Note that we can write $e^{\calL t}$ as the convex combination
			\begin{align*}
			e^{\calL t}
			&= \sum_{k=0}^\infty \frac{t^k}{k!} (\lambda\numkac(Q-\bbone) + \mu\numkac)(R-\bbone))^k \\
			&= e^{-(\lambda+\mu)\numkac t} \sum_{k=0}^\infty \frac{((\lambda+\mu)\numkac t)^k}{k!} \left( \frac{\lambda}{\lambda+\mu} Q + \frac{\mu}{\lambda+\mu} R \right)^k.
			\numberthis
			\end{align*}
			Since $||Q||\leq 1$ and $||R||\leq 1$ by the previous parts, we get
			\begin{equation}
			\left|\left| \frac{\lambda}{\lambda+\mu} Q + \frac{\mu}{\lambda+\mu} R \right|\right|
			\leq \frac{\lambda}{\lambda+\mu} ||Q|| + \frac{\mu}{\lambda+\mu} ||R||
			\leq 1.
			\end{equation}
			This proves that
			\begin{equation}
			\left| \left| e^{\calL t} \right| \right|
			\leq 1.
			\end{equation}
			
			\item Let $f_0 \in \LOneKac$. Then, we get that similar to part \ref{RemKacThermoRegParti} and \ref{RemKacThermoRegPartii} that
			\begin{equation}
			\intrtN Qf_0 \,\textup{d}\boldv
			= \intrtN f_0 \,\textup{d}\boldv, \quad
			\intrtN R_j f_0 \,\textup{d}\boldv
			= \intrtN f_0 \,\textup{d}\boldv.
			\end{equation}
			Termwise integration yields
			\begin{align*}
			\intrtN e^{\calL t}f_0 \,\textup{d}\boldv
			&= \sum_{k=0}^\infty \frac{t^k}{k!} \intrtN (\lambda\numkac(Q-\bbone) + \mu\numkac(R-\bbone))^k f_0 \,\textup{d}\boldv \\
			&= \intrtN (\lambda\numkac(Q-\bbone) + \mu\numkac(R-\bbone))^0 f_0 \,\textup{d}\boldv \\
			&= \intrtN f_0 \,\textup{d}\boldv.
			\numberthis
			\end{align*}
			Note that we can integrate term by term since $\calL$ is bounded, and therefore the exponential series converges uniformly.
		\end{enumerate}
	\end{rem}

	\begin{rem}[Regularity of the Kac system coupled to a heat reservoir]
		\hfill
		
		\begin{enumerate}[label=\roman*)]
			\item The time evolution operator $\calL$ generates a contraction semigroup. This is precisely what provides us the following majorant needed to exchange the order of integration and differentiation in the proofs of this chapter. For all $t\geq 0$ and $\boldu\in\bbR^{\di(\numkac+\numhr)}$ we get
			\begin{equation}
			\left|\frac{\textup{d}}{\textup{d}t} F_t(\boldu)\right|
			= \left| \calL e^{\calL t} F_0(\boldu) \right|
			\leq ||\calL||\, \left|\left|e^{\calL t}\right|\right|\, |F_0(\boldu)|
			\leq ||\calL||\, |F_0(\boldu)|
			\end{equation}
			
			\begin{proof}
				The claim follows from the decomposition \eqref{EqColHis} of $e^{\calL t}$ in collision histories. Since $||R_\alpha||\leq 1$ for all $\alpha \in I$, we get $||R^{\balpha}||\leq 1$ for all $\balpha \in I^k$. This yields
				\begin{equation}
				\left|\left|e^{\calL t}\right|\right|
				\leq e^{-\Lambda t} \sum_{k=0}^\infty \frac{(\Lambda t)^k}{k!} \sum_{\balpha \in I^k} \lambda^{\balpha} ||R^{\balpha}||
				\leq 1.
				\end{equation}
			\end{proof}
			
			\item For a solution $(F_t)_{t\geq0}\in\LOneKNM$ of the Kac master equation \eqref{EqMasterEqHeatRes} the following statements are equivalent:
			\begin{enumerate}[label=\alph*)]
				\item \label{RemBasicRegHRPart1}
				$F_0\in\LOneKNM$ is a PDF.
				\item \label{RemBasicRegHRPart2}
				$F_t\in\LOneKNM$ is a PDF for all $t\geq0$.
				\item \label{RemBasicRegHRPart3}
				$f_t\in\LOneKac$ is a PDF for all $t\geq0$.
				\item \label{RemBasicRegHRPart4}
				$h_t\in\LOneGauss$ is a PDF with respect to the Gaussian measure $\MwN(\boldv)\textup{d}\boldv$.
			\end{enumerate}
			
			\begin{proof}
				The equivalence of \ref{RemBasicRegHRPart3} and \ref{RemBasicRegHRPart4} is clear by the definition \eqref{EqDefh_t}. Since by definition \eqref{EqDeff_t}
				\begin{equation}
				\intrtN f_t \,\textup{d}\boldv
				= \intrtNM F_t \,\textup{d}\boldu
				\end{equation}
				the statements \ref{RemBasicRegHRPart2} and \ref{RemBasicRegHRPart3} are equivalent as well.
				
				Using the representation \eqref{EqColHis} of $e^{\calL t}$ in terms of collision histories we get the equivalence of the first two statements by
				\begin{align*}
				&\intrtNM F_t(\boldu) \,\textup{d}\boldu
				= e^{-\Lambda t} \sum_{k=0}^\infty \frac{(\Lambda t)^k}{k!} \sum_{\balpha\in I^k} \lambda^{\balpha} \intrtNM \textup{d}\boldu \, Q^k[F_0](\boldu) \\
				&= e^{-\Lambda t} \sum_{k=0}^\infty \frac{(\Lambda t)^k}{k!} \sum_{\balpha\in I^k} \lambda^{\balpha} \intkbbS \textup{d}\rho(\bsigma) \intrtNM \textup{d}\boldu\,
				F_0\left( \boldM_{\sigma_k}^{(\alpha_k)} \cdots \boldM_{\sigma_1}^{(\alpha_1)} \boldu \right) \\
				&= e^{-\Lambda t} \sum_{k=0}^\infty \frac{(\Lambda t)^k}{k!} \sum_{\balpha\in I^k} \lambda^{\balpha} \intkbbS \textup{d}\rho(\bsigma) \intrtNM \textup{d}\boldp\,
				F_0\left( \boldp \right) \\
				&= \intrtNM F_0\left( \boldp \right) \,\textup{d}\boldp,
				\numberthis
				\end{align*}
				where we used the transformation $\boldp \coloneqq \boldM_{\sigma_k}^{(\alpha_k)} \cdots \boldM_{\sigma_1}^{(\alpha_1)} \boldu$.
			\end{proof}
		\end{enumerate}
	\end{rem}

	\section{Ornstein-Uhlenbeck semigroup}
	\label{SecOrnsteinUhlenbeckSemigroup}

	In this chapter, we summarize the relevant properties of the Ornstein-Uhlenbeck semigroup and provide proofs as well. Hereby, we always consider the Gaussian measure $\MwN(v)\textup{d}v$, where
	\begin{equation}
	\label{EqOUGM}
	\MwN(v)
	\coloneqq \left(\frac{\invtemp}{2\pi}\right)^\frac{\di}{2} e^{-\frac{\invtemp}{2}v^2}.
	\end{equation}
	
	\begin{defi}
		\label{DefiOUsemigroup}
		The \textit{Ornstein-Uhlenbeck semigroup} is defined by (see \cite{LedouxIntCrit}, p. 444)
		\begin{equation}
		\label{EqDefOUsemigroup}
		P_s\testH(v)
		\coloneqq \intrt \testH\left(e^{-s}v + \sqrt{1-e^{-2s}} x \right) \,\MwS(x)\textup{d}x,
		\quad s \in [0,\infty), v \in \bbR^\di,
		\end{equation}
		where $P_s$ acts on $L^1(\bbR^\di, \MwS(x)\textup{d}x)$.
	\end{defi}
	
	\begin{lem}
		The Ornstein-Uhlenbeck semigroup is a strongly continuous semigroup.
		\begin{proof}
			It is clear that $P_0 = \bbone$, since
			\begin{equation}
			P_0\TS(v)
			= \int \TS(v) \,\MwS(x)\textup{d}x
			= \TS(v).
			\end{equation}
			Further, we have by the dominated convergence theorem that
			\begin{align*}
			\lim_{s\to 0}P_s\TS(v)
			&= \lim_{s\to 0} \int \TS\left( e^{-s}v + \sqrt{1-e^{-2s}}x \right) \,\MwS(x)\textup{d}x \\
			&= \int \lim_{s\to 0} \TS\left( e^{-s}v + \sqrt{1-e^{-2s}}x \right) \,\MwS(x)\textup{d}x
			= \TS(v).
			\numberthis
			\end{align*}
			This proves, that $\slim_{s\to0} P_s = \bbone$. Finally, we verify that $P_{s+t} = P_sP_t$ for all $s,t \geq 0$. We observe that
			\begin{align*}
			P_sP_t&[\TS](v)
			= \intrt \MwS(x)\textup{d}x \intrt \MwS(y)\textup{d}y\,
			\TS\left( e^{-t}\left(e^{-s}v + \sqrt{1-e^{-2s}}x \right) + \sqrt{1-e^{-2t}}y \right) \\
			&= \intrt \MwS(x)\textup{d}x \intrt \MwS(y)\textup{d}y\,
			\TS\left( e^{-(s+t)}v + e^{-t}\sqrt{1-e^{-2s}}x + \sqrt{1-e^{-2t}}y \right).
			\label{EqProofPsPtStart}\numberthis
			\end{align*}
			Defining
			\begin{equation}
			\begin{split}
			a
			&\coloneqq \frac{e^{-t}\sqrt{1-e^{2s}}}{\sqrt{1-e^{-2(s+t)}}}, \\
			p
			&\coloneqq a x + b y,
			\end{split}
			\quad
			\begin{split}
			b
			&\coloneqq \frac{\sqrt{1-e^{-2t}}}{\sqrt{1-e^{-2(s+t)}}}, \\
			q
			&\coloneqq - b x + a y.
			\end{split}
			\end{equation}
			yields
			\begin{equation}
			\begin{split}
			a^2 + b ^2
			&= \frac{e^{-2t}\left(1-e^{-2s}\right) + 1 - e^{-2t}}{1-e^{-2(s+t)}}
			= \frac{1-e^{-2(s+t)}}{1-e^{-2(s+t)}}
			= 1, \\
			p^2 + q^2
			&= \left(a^2+b^2\right)\left(x^2+y^2\right)
			= x^2 + y^2.
			\end{split}
			\end{equation}
			Therefore, the transformation $\phi(x,y) \coloneqq (p,q)$ is a rotation with
			\begin{equation}
			\begin{split}
			\det(\phi'(x,y))
			&= \det\left(\begin{pmatrix} a & -b \\ b & a \end{pmatrix}\right)
			= a^2 + b^2
			= 1, \\
			\MwS(x)\MwS(y)
			&= \MwS(p)\MwS(q).
			\end{split}
			\end{equation}
			Hence, equation \eqref{EqProofPsPtStart} transforms into
			\begin{align*}
			P_sP_t[\TS](v)
			&= \intrt \MwS(p)\textup{d}p \intrt \MwS(q)\textup{d}q\,
			\TS\left( e^{-(s+t)}v + \sqrt{1-e^{-2(s+t)}} p \right) \\
			&= \intrt \MwS(p)\textup{d}p\,
			\TS\left( e^{-(s+t)}v + \sqrt{1-e^{-2(s+t)}} p \right) \\
			&= P_{s+t}[\TS](v).
			\numberthis
			\end{align*}
		\end{proof}
	\end{lem}
	
	\begin{lem}
		The Ornstein-Uhlenbeck semigroup is self-adjoint (see \cite{LedouxIntCrit}), i.e.\ $P_s$ is self-adjoint on $L^2(\bbR^\di,\MwS(x)\textup{d}x)$ for all $s\in[0,\infty)$. In particular, it follows since $P_s[1] = 1$ that
		\begin{equation}
		\intrt P_s\TS \,\MwS(x)\textup{d}x
		= \intrt \TS \,\MwS(x)\textup{d}x.
		\end{equation}
		
		\begin{proof}
			For all functions $F,G \in L^2(\bbR^\di,\textup{d}\MwS(v)\textup{d}v)$, we have
			\begin{align*}
			\lscal P_sF, G \rscal
			&= \intrt \textup{d}v\, \MwS(v) P_sF(v) G(v) \\
			&= \intrt \intrt \textup{d}v\,\textup{d}w\, \MwS(v)\MwS(w) F\left(e^{-s} v + \sqrt{1-e^{-2s}} w\right) G(v).
			\numberthis
			\end{align*}
			Defining 
			\begin{equation}
			\begin{split}
			p &\coloneqq e^{-s} v + \sqrt{1-e^{-2s}} w \\
			q &\coloneqq - \sqrt{1-e^{-2s}} u + e^{-s} w
			\end{split}
			\quad\Leftrightarrow\quad
			\begin{split}
			v &= e^{-s} p - \sqrt{1-e^{-2s}} q \\
			w &= \sqrt{1-e^{-2s}} p + e^{-s} q,
			\end{split}
			\end{equation}
			yields the transformation $\phi(v,w) \coloneqq (p,q)$ with
			\begin{equation}
			\begin{aligned}
			\left(e^{-s}\right)^2 + \left(\sqrt{1-e^{-2s}}\right)^2 &= 1,
			&\det(\phi'(u,w)) &= 1, \\
			p^2 + q^2 &= v^2 + w^2,
			&\MwS(u)\MwS(w) &= \MwS(p)\MwS(q).
			\end{aligned}
			\end{equation}
			Therefore, the claim follows by
			\begin{align*}
			\lscal P_sF, G \rscal
			&= \intrt \intrt \textup{d}p\,\textup{d}q\, \MwS(p)\MwS(q) F(p) G\left(e^{-s} p - \sqrt{1-e^{-2s}} q\right) \\
			&= \intrt \intrt \textup{d}p\,\textup{d}q\, \MwS(p)\MwS(-q) F(p) G\left(e^{-s} p + \sqrt{1-e^{-2s}} q\right) \\
			&= \lscal F, P_sG \rscal,
			\numberthis
			\end{align*}
			where we used in the last equality, that the Gaussian $\MwS$ is centered, i.e.\ $\MwS(-q) = \MwS(q)$.			
		\end{proof}
	\end{lem}
	
	Next, we investigate the generator of the Ornstein-Uhlenbeck semigroup. In \cite[Lemma 13.1.1]{InfDimAna} an explicit formula is given for the generator:
	
	\begin{lem}
		The generator $L$ of the Ornstein-Uhlenbeck semigroup is given by
		\begin{equation}
		L\testH(v)
		= \frac{1}{\invtemp} \Delta \TS(v) - v \cdot \nabla\TS(v),
		\end{equation}
		where we recall that $\invtemp$ is the inverse temperatur of the Gaussian $\MwN$ as given in \eqref{EqOUGM}.
		
		\begin{proof}
			The generator $L$ is for any test function $\TS \in D(L)$ in its domain given by
			\begin{equation}
			L[\TS]
			= \lim_{t \to 0} \frac{P_t\TS-\TS}{t}
			= P_s \lim_{t \to 0} \frac{P_t\TS-\TS}{t}\bigg|_{s=0}
			= \lim_{t \to 0} \frac{P_{s+t}\TS-P_s\TS}{t}\bigg|_{s=0}
			= \frac{\textup{d}}{\textup{d}s} P_s\TS\bigg|_{s=0}.
			\end{equation}
			Hence, we compute the derivation of $P_s\TS$ with respect to $s$. For shorter notation, we define $\OUalpha(s) \coloneqq e^{-s}$ and $\OUbeta(s) \coloneqq \sqrt{1-e^{-2s}}$. Since we do not know the derivation of $\TS$, but the derivation of $\MwS$, we use the transformation $p \coloneqq \OUalpha v + \OUbeta w \Leftrightarrow w = \frac{p-\OUalpha v}{\OUbeta}$ to get
			\begin{equation}
			\label{EqProofGeneratorTrans}
			P_s\TS(v)
			= \frac{1}{\OUbeta^\di} \intrt\textup{d}p\, \MwS\left( \frac{p-\OUalpha v}{\OUbeta} \right) \TS(p).
			\end{equation}
			Now, we can compute
			\begin{equation}
				\label{EqProofGeneratorddsPsG}
				\frac{\textup{d}}{\textup{d}s} P_s\TS(v)
				= \intrt\textup{d}w\, \MwS(w) \TS(\OUalpha v + \OUbeta w) \left( -\di\frac{\OUalpha^2}{\OUbeta^2} - \invtemp\frac{\OUalpha}{\OUbeta} v \cdot w + \invtemp\frac{\OUalpha^2}{\OUbeta^2} w^2 \right),
			\end{equation}
			A lengthy computation yields
			\begin{equation}
			\label{EqProofGeneratorNabla}
			\nabla_v P_s\TS(v)
			= \intrt \textup{d}w\, \MwS(w) \TS(\OUalpha v + \OUbeta w) \cdot \invtemp \frac{\OUalpha}{\OUbeta} w
			\end{equation}
			and
			\begin{equation}
				\label{EqProofGeneratorLaplace}
				\Delta_v P_s\TS(v)
				= \intrt \MwS(w)\textup{d}w\, \TS(\OUalpha v + \OUbeta w) \left(\invtemp^2 \frac{\OUalpha^2}{\OUbeta^2} w^2 -\invtemp\di \frac{\OUalpha^2}{\OUbeta^2}\right).
			\end{equation}
			Combining \eqref{EqProofGeneratorddsPsG}, \eqref{EqProofGeneratorNabla}, and \eqref{EqProofGeneratorLaplace} we get
			\begin{equation}
			\frac{\textup{d}}{\textup{d}s} P_s\TS(v)
			= \frac{1}{\invtemp} \Delta_v P_s\TS(v) - v \cdot \nabla_v P_s\TS(v).
			\end{equation}
			Evaluating at $s = 0$ completes the proof.
		\end{proof}
	\end{lem}

	\bibliographystyle{plain}
	\bibliography{references}

\end{document}